\documentclass[journal]{IEEEtran}
\usepackage[utf8]{inputenc}
\usepackage[T1]{fontenc}
\usepackage{algorithm}
\usepackage{amsmath}
\usepackage{amsthm}
\usepackage{amssymb}
\usepackage{graphicx}
\usepackage{setspace}
\usepackage{epstopdf}
\usepackage[font={normal}]{caption}
\usepackage{cite}
\usepackage{bm}
\usepackage{algpseudocode}
\usepackage{balance}
\usepackage{array,multirow,makecell}\setcellgapes{1pt}
\usepackage{comment}
\usepackage[table]{xcolor}
\usepackage{setspace}
\usepackage{subfigure}

\DeclareMathOperator*{\argmin}{argmin}
\newcolumntype{R}[1]{>{\raggedleft\arraybackslash }b{#1}}
\newcolumntype{L}[1]{>{\raggedright\arraybackslash }b{#1}}
\newcolumntype{C}[1]{>{\centering\arraybackslash }b{#1}}
\makegapedcells

\begin{document}

\title{Optimal Resource Management for Hierarchical Federated Learning over HetNets with Wireless Energy Transfer}
\author{Rami~Hamdi,~\IEEEmembership{Member,~IEEE,}
	Ahmed~Ben~Said,
        Emna~Baccour,
	Aiman~Erbad,~\IEEEmembership{Senior~Member,~IEEE,}	
    Amr~Mohamed,~\IEEEmembership{Senior~Member,~IEEE,}
	Mounir~Hamdi,~\IEEEmembership{Fellow,~IEEE,}
	and~Mohsen~Guizani,~\IEEEmembership{Fellow,~IEEE}
\thanks{A preliminary version of this work~\cite{gc} has appeared in the proceedings of IEEE Global Communications Conference in 2021.} 
\thanks{R. Hamdi is with the AFG College with the University of Aberdeen, Doha, Qatar (email: rami.hamdi@afg-aberdeen.edu.qa).}
\thanks{A. B. Said and A. Mohamed are with the College of Engineering, Qatar University, Doha, Qatar (email: abensaid@qu.edu.qa; amrm@qu.edu.qa).}
\thanks{E. Baccour, A. Erbad, and M. Hamdi are with the Division of Information and Computing Technology, College of Science and Engineering, Hamad Bin Khalifa University, Qatar Foundation, Doha, Qatar (email: ebaccourepbesaid@hbku.edu.qa; aerbad@ieee.org; mhamdi@hbku.edu.qa).}
\thanks{M. Guizani is with  the Machine Learning Department, Mohamed Bin Zayed University of Artificial Intelligence (MBZUAI), Abu Dhabi, UAE (email: mguizani@ieee.org).}
}
\maketitle
\thispagestyle{empty}

\begin{abstract}
Remote monitoring systems analyze the environment dynamics in different smart industrial applications, such as occupational health and safety, and environmental monitoring. Specifically, in industrial Internet of Things (IoT) systems, the huge number of devices and the expected performance put pressure on resources, such as computational, network, and device energy. Distributed training of Machine and Deep Learning (ML/DL) models for intelligent industrial IoT applications is very challenging for resource limited devices over heterogeneous wireless networks (HetNets). Hierarchical Federated Learning (HFL) performs training at multiple layers offloading the tasks to nearby Multi-Access Edge Computing (MEC) units. In this paper, we propose a novel energy-efficient HFL framework enabled by Wireless Energy Transfer (WET) and designed for heterogeneous networks with massive Multiple-Input Multiple-Output (MIMO) wireless backhaul. Our energy-efficiency approach is formulated as a Mixed-Integer Non-Linear Programming (MINLP) problem, where we optimize the HFL device association and manage the wireless transmitted energy. However due to its high complexity, we design a Heuristic Resource Management Algorithm, namely H2RMA, that respects energy, channel quality, and accuracy constraints, while presenting a low computational complexity. We also improve the energy consumption of the network using an efficient device scheduling scheme. Finally, we investigate device mobility and its impact on the HFL performance. Our extensive experiments confirm the high performance of the proposed resource management approach in HFL over HetNets, in terms of training loss and grid energy costs. 

\end{abstract}

\section{Introduction}
Occupational health and safety (OHS) has recently attracted the attention of governmental agencies in order to provide workers' protection and minimize health risks~\cite{oc1,oc2,oc3}. More generally, remote monitoring systems that capture the environment dynamics and monitor individuals are used in different applications to help senior citizens, patients, and factory workers. Distributed Artificial Intelligence (AI)~\cite{ai} and communication technologies allow continuous monitoring for Key Health Indicators (KHIs) of a large number of workers using relevant industrial IoT devices (smart phones, mobile devices, wearables, smart cameras, and smart watches). The collected KHIs are then transmitted to a central or a nearby edge Central Unit (CU) to be processed by machine and deep learning algorithms and help in the decision making tasks. 

With an increasing number of connected devices, centralized Machine and Deep Learning (ML/DL) algorithms require more and more memory resources and huge processing (i.e. CPU)~\cite{fl0}.
In a wireless network setting with real-time and low-latency requirements, a centralized learning scheme becomes inefficient as it induces delay and may compromise the privacy of exchanged data between devices. Hence, decentralized learning approaches have emerged, particularly Federated Learning (FL)~\cite{fl0,fl1}. FL enables to train a shared model among multiple devices, where each device uses only its private local data~\cite{fl1}. Indeed, the devices are training and updating the weights of the global model they received using their local data samples. Then, they send the new model weights to a CU for aggregation into a shared learning model. Finally, the CU broadcasts the aggregated model weights to the FL participants in order to update their models. This routine is iterated until reaching the desired model performance. FL addresses the challenge of data privacy, while enhancing the bandwidth usage as the amount of transferred data to the CU decreases. However, with the nexus of thousands of devices, several challenges need to be addressed including communication bottlenecks, increase in energy consumption, congestion, and and data skewness i.e. non-i.i.d data~\cite{hfl4}.
To cope with these issues, Hierarchical Federated Learning (HFL) has been presented~\cite{hfl1,hfl2}.
HFL enables offloading the computation task to the Multi-Access Edge Computing (MEC) in device vicinity, hence increasing the energy efficiency and enhancing the  response time  of the IoT nodes. In this new configuration, the devices send their local updates to their corresponding MEC for learning and model aggregation. Then, the MECs transfer their aggregated models to the CU in order to compute the global aggregation. The most distinctive feature of HFL is the ability to address and manage diverse data distribution, e.g., non-i.i.d property of the data. This is achieved by carefully assigning devices to MECs, hence speeding up convergence and attaining higher performance levels~\cite{hfl4}. We highlight that if the data is i.i.d, the HFL may not have a big contribution to the system performance.

In a wireless communication environment with constant inter-device interference, latency constraint, increase in energy consumption and device heterogeneity, the implementation of HFL becomes challenging. Therefore, many research efforts have recently focused on making HFL practical for wireless communication networks. More precisely, recent works have tackled resources allocation problems, including wireless resource allocation, edge-device assignment,  device selection and energy management with the objective of optimizing the performance of such system fully exploiting its potential. On the other hand, Wireless Energy Transfer (WET) represents a promising technology whose objective is to address the problem of energy bottlenecks in wireless networks~\cite{wet}. The system is based on using electromagnetic radiation waves transmitted from a power source to deliver energy to a wireless device. Delivering power using wireless technology satisfies the frequent charging for mobile devices. Hence, it represents a key technology for the design of energy-efficient IoT wireless networks~\cite{wet2}.

One of the challenges facing the ML/DL distribution is the high-energy consumption. Thus, to enhance the energy efficiency of HFL processed over HetNets with wirless backhaul, we propose to incorporate energy harvesting~\cite{tvt17}, where each MEC transmits the harvested energy to its related devices via WET.  Compared to the existing schemes in~\cite{hfl1,hfl2,hfl4, hfl3,hfl5,hfl6,hfl7,hfl8,hfl9,hfl10,hfl11,hfl12,hfl13,hfl14,hfl15,hfl16}, our work is distinctive as it focuses on designing an energy-efficient HFL architecture assisted by WET.
As per our knowledge, we are the first to consider the implementation of FL over HetNets with massive MIMO wireless backhaul enabled by WET. Furthermore, a practical mobility scenario that represents the workers in a construction site is considered by developing an efficient dynamic resource allocation scheme. The contributions of our work include the following:
\begin{itemize}
\item We propose a novel energy-efficient HFL framework, designed for HetNets with massive MIMO wireless backhaul and enabled by WET.
\item We formulate our energy-efficiency approach as a Mixed-Integer Non-Linear Programming, where we optimize the HFL device association and manage the wireless transmitted energy.
\item We derive the optimal device association analytically and we use Brute-Force search to obtain the energy management decisions.
\item Due to the high complexity of the optimal solution, we propose a device association and energy management algorithm that aims to enhance the energy-efficiency of the proposed HFL framework.
\item Some devices with high energy requirements may be deactivated for some frames. Hence, we propose a device scheduling strategy to decrease the energy consumption without deteriorating the learning accuracy.
\item We develop an efficient dynamic device association scheme, where we consider a mobility scenario that represents the workers in a construction site.	
\item Finally, we conduct an extensive evaluation of the proposed system to prove the energy efficiency and high accuracy performance of our resource management, device association, and device scheduling schemes under different network configurations.
\end{itemize}

We organize the remainder of this article as follows: In section II, we explore the related works that study one-layer and hierarchical FL. In section III, we present the system model. The optimization problem is formulated in Section IV. We investigate the resource management in Section V. The impact of mobility in FL performance is investigated in Section VI. Evaluation results are provided in Section VII. Lastly, conclusions are drawn in Section VIII. 

\begin{table*}[t]
	\centering
	\caption{List of key Notations.}
	\label{tab1}
	\begin{tabular}{|C{1.5cm}|L{5.5cm}||C{1.5cm}|L{5.5cm}|}
		\hline   \textbf{Symbol} & \textbf{Description} & \textbf{Symbol} & \textbf{Description}\\
		\hline   $N_{\text{cu}}$ & Number of antennas at the CU  & $\mathbf{H}_m$ & Channel matrix between the CU and MEC $m$\\
		\hline   $N_{\text{mec}}$ & Number of antennas at each MEC  &  $\mathbf{g}_{m,k}$ & Channel vector between device $k$ and MEC $m$ \\
		\hline   $K$ & Number of devices  & $\beta_{m,k}$  & Pathloss \\
		\hline   $L$ & Number of frames  & $\sigma^2$ & Noise variance\\
		\hline   $M$ &  Number of MECs  & $\mathbf{W}_m$  & Detection matrix at MEC $m$\\
		\hline   $T$ & Duration of each frame  & $E^{\text{dev}}_{m,k}(l)$ & Energy required by  device $k$ assigned to MEC $m$ at frame $l$\\
		\hline   $\Lambda_m$  & Set of devices associated with MEC $m$  & $E^{\text{cir}}$ & Fixed circuit energy consumption \\
		\hline   $K_m$ & Cardinal of $\Lambda_m$   & $E^{\text{ac}}_{m,k}(l)$ & Required transmit energy of device $k$ at frame $l$\\
		\hline   $C$ & Number of classes  & $Q$ & Size of the local FL model\\
		\hline   $\mathcal{D}_k$ & Dataset of device $k$  & $B_{max}$ & Maximal battery capacity\\
		\hline   $\boldsymbol{x}_{k}$ & Input of device $k$ & $A_{m,k}(l)$ & Amount of harvested energy by device $k$ assigned to MEC $m$ at frame $l$\\
		\hline   $\boldsymbol{y}_k$ & Output of device $k$ & $B_k(l)$ & Battery level of device $k$ at frame $l$ \\
		\hline   $\boldsymbol{S}$ & Data matrix of the devices  & $E^{\text{wet}}_m(l)$  & Transferred energy from MEC $m$ at frame $l$ \\
		\hline   $S_{c,k}$ & Number of training data samples of the device $k$ for the class $c$  & $\alpha_{m}(l)$ & Grid's weights \\
		\hline   $\boldsymbol{w}_k(l)$ & Weight vector of the local ML model of device $k$ at frame $l$  & $\Delta$ & Total grid energy consumption cost \\
		\hline   $\cup_{j \in \Lambda_m} \mathcal{D}_j$ & Virtual dataset available at MEC $m$  & $\chi_{m,k}$  & Device association index \\
		\hline   $p^{(m)}(i)$ & Probability of the class $i$ per MEC $m$’s distribution  &  $p(\cdot)$ & Global classes distribution \\
		\hline 
	\end{tabular}
\end{table*}

\section{Related Work}
\subsection{Federated Learning}
A number of studies have explored resource allocation for FL~\cite{fl_rami,fl2,fl3,fl4,fl5,fl6,fl7,wet_fl1,wet_fl2,wet_fl3,wet_fl4} over wireless networks. In~\cite{fl_rami}, an energy harvesting FL framework was developed by proposing efficient user association and scheduling schemes. Minimizing the energy costs of FL wireless systems was investigated, while taking into consideration a limited energy budget by developing an efficient resource management scheme. In~\cite{fl3}, a data augmentation and  adaptive user scheduling scheme was developed to overcome the challenges related to the imbalanced distribution of data in FL. The convergence time was theoretically investigated and optimized in~\cite{fl4} by designing an efficient control scheme. The FL performance was improved in~\cite{fl5} by proposing an efficient user scheduling policy. Moreover, the authors of~\cite{fl6} designed an optimal user selection algorithm that minimizes the FL convergence rate. In~\cite{fl7}, a user selection scheme was proposed to decrease the effect of unreliable users and enhance the accuracy of federated learning training. To address the issue of battery limitation for FL users, initial works~\cite{wet_fl1,wet_fl2,wet_fl3,wet_fl4} incorporated WET systems when performing FL tasks. In~\cite{wet_fl1}, the authors assumed that the unmanned aerial vehicles (UAVs) have WET capability and proposed a suitable FL solution in terms of energy efficiency. In~\cite{wet_fl2}, a heterogeneous mobile architecture is implemented and an optimal resource allocation solution is proposed that minimizes the energy consumption of FL devices. In~\cite{wet_fl3}, a low-complexity beamforming strategy was proposed to minimize the mean-square error in FL based systems enabled by WET. The authors of~\cite{wet_fl4} developed an analytic FL framework that establishes a trade-off between the model convergence and the power consumption which represents a guideline to ensure efficient learning tasks.

\subsection{Hierarchical Federated Learning}
Others works have tackled the optimization of resource allocation for HFL~\cite{hfl2,hfl4,hfl9,hfl3,hfl7,hfl1,hfl5,hfl11,hfl8,hfl12,hfl13,hfl14,hfl15,hfl16,hfl6,hfl10}. More specifically, the authors in~\cite{hfl2} used HFL to design a reliable anomaly detection strategy for IoT systems. Moreover, to face the challenges caused by non-i.i.d. data, the authors in~\cite{hfl4} designed an edge-device assignment scheme to enhance the federated averaging performance. Considering the same constraint of non-i.i.d. data among users, the authors in~\cite{hfl9} proposed an efficient resource allocation and user assignment scheme in HFL for heterogeneous IoT systems. Meanwhile, the authors of~\cite{hfl3} proposed a novel HFL architecture aiming to minimize the consumed energy and the total training latency under challenging data distribution. Energy-efficiency was also the focus of the work in~\cite{hfl7}, where authors have introduced a power control and client association solution for HFL. The work in~\cite{hfl1} targeted the minimization of the global cost using a novel resource allocation and device association algorithm. Another extremely important metric in FL systems is the latency, that was the objective of the works in~\cite{hfl11} and~\cite{hfl5}. While the former optimized the association between multiple servers and devices, the latter designed a HFL framework across heterogeneous cellular networks. The authors in~\cite{hfl8} proposed a flexible decentralized control over the training process in order to enhance the privacy of HFL, while the authors of~\cite{hfl12} suggested using blockchain to meet the privacy requirements of internet of vehicles composing the HFL system.
In~\cite{hfl13}, a novel HFL framework based on clustering is proposed to enhance the test accuracy. The authors of~\cite{hfl14} investigated HFL by proposing a hybrid data partitioning algorithm. In~\cite{hfl15}, an efficient HFL algorithm is presented with an objective to determine the optimal cluster structure. In~\cite{hfl16}, a resource allocation heuristic was designed to guarantee the HFL convergence rate subject to fairness constraint. Finally, the game theory was used in~\cite{hfl6} and~\cite{hfl10} to design an efficient resource allocation strategy for HFL.

Unlike the works proposed in the literature, we present in this paper a novel HFL architecture considering massive MIMO wireless backhaul, i.e. the links between the MECs and the devices and the links between MECs are wireless. Moreover, we consider a practical mobility scenario that represents the workers in a construction site. Additionally, the use of energy harvesting for HFL was not considered in most of the related works. Hence, we consider in this work that the MECs have WET capability and the devices are equipped with energy harvesting batteries, which allows to enhance the energy efficiency. Taking into account these challenging constraints, we formulate and investigate a novel resource optimization problem.

\section{System Model and Energy Model}
\subsection{Hierarchical Federated Learning Model}
\begin{figure}[t]
	\centerline{\includegraphics [width=1\columnwidth]  {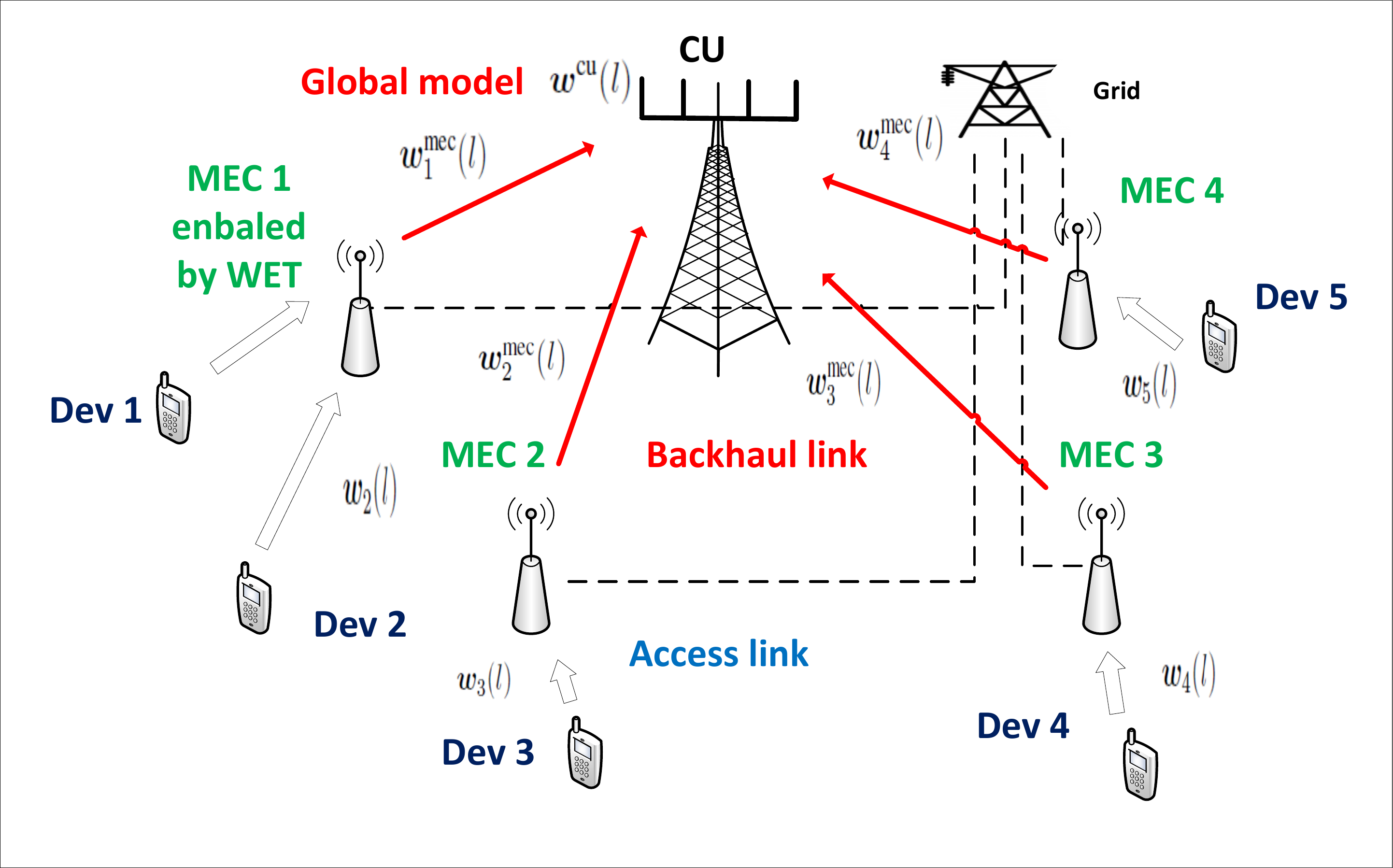}}
	\caption{WET-enabled HFL over HetNets in battery powered end-devices.}
	\label{fig1}
\end{figure}

Massive MIMO is based on multiplexing a large number of antennas simultaneously to satisfy the demand of  multiple users in the same time-frequency resource, which contributes to improving  energy efficiency, spectral efficiency, and latency~\cite{mimo}. Furthermore, HetNets, which are based on network densification in multiple dimensions with different types of cells, allow to efficiently exploit the wireless resources and improve the QoS. Incorporating massive MIMO and HetNets in IoT wireless networks is suitable to optimize the FL performance. Hence, we opt for a HetNet as presented in Fig.~\ref{fig1}. Our system is composed of a massive MIMO Central Unit (CU) equipped with $N_{\text{cu}}$ antennas, and $M$ MECs, each one is endowed with $N_{\text{mec}}$  MIMO antennas ($N_{\text{cu}} > M N_{\text{mec}} $). Different MECs communicate with the CU through a wireless backhaul link~\cite{wb}. Furthermore, each MEC $m$ serves multiple single-antenna devices. Let $\Lambda_m$ be the set of $K_m$ devices served by the MEC $m$. We note that the total number of devices in the network is equal to $K$. We study our system over an interval of time that we divide into $L$ frames of duration $T$.  
Table~\ref{tab1} presents different key notations introduced in our paper. The device association index is defined as:
\begin{equation}
	\label{eq:0}
	\chi_{m,k}=\left\{\begin{array}{ll}1, & \text{if device } k \text{ is associated to } \text{MEC}~m, \\ 0, & \text{otherwise}.
	\end{array}\right.
\end{equation}
Each device $k$ owns a dataset $\mathcal{D}_k=\{\boldsymbol{x}_k,\boldsymbol{y}_k\}$, where $\boldsymbol{x}_{k}$ and $\boldsymbol{y}_k$ denote the input and output of the device $k$. These devices train their FL models locally and send the obtained weights to their related MECs. Then, each MEC aggregates the weights of its devices and forwards the results to the CU. Finally, the CU aggregates the weights received from all MECs and shares the updated FL model with all devices. This process is repeated until reaching a desirable accuracy. Let the Matrix $\boldsymbol{S} \in \mathbb{N}^{C\times K}$ describe the splitting of data among all devices, where $C$ denotes the total number of classes in the learning problem. Particularly, each element $S_{c,k}$ represents the number of data samples owned by the device $k$ and belonging to the class $c$. 

Let $\boldsymbol{w}_k(l)$ represent the vector of FL weights of the local model trained within the device $k$ at a frame $l$. After aggregating all weights, first incoming from the devices, then collected from the MECs, the CU is considered globally fitting the resultant weights by minimizing a particular loss function. In our work, we opt for the cross-entropy loss function that can be presented as follows:
\begin{equation}
	\label{eq:1}
	\mathcal{L}\left(\boldsymbol{w}\right) =\sum_{c=1}^{C} -p(c) \mathbb{E}_{\boldsymbol{x} \sim q(\boldsymbol{x} \mid y=c)}\left[\log d_{c}(\boldsymbol{x} ; \boldsymbol{w})\right],
\end{equation}

where $p(\cdot)$ defines the global distribution of classes, $q(\cdot \mid \cdot)$ denotes the likelihood function, and $d_{c}(\boldsymbol{x} ; \boldsymbol{w})$ represents the probability that the input $\boldsymbol{x}$ belongs to the class $c$, under the parameters $\boldsymbol{w}$.\\ 
After receiving the weights of the local FL models from the related devices, each MEC computes the aggregation following the equation in~\cite{hfl4}:  
\begin{equation}
	\label{eq:2}
	\boldsymbol{w}^{\text{mec}}_m(l)= \sum_{k \in \Lambda_m} \frac{ \mid \mathcal{D}_k \mid }{  |\cup_{j \in \Lambda_m} \mathcal{D}_j| }  \boldsymbol{w}_k(l),
\end{equation}
where $\cup_{j \in \Lambda_m} \mathcal{D}_j$ is the grouping of local datasets owned by different devices related to the MEC $m$. Accordingly, it can be considered as a virtual dataset trained at the MEC $m$. The next step is to aggregate the weights received from all the MECs at the central unit, as done in~\cite{hfl4}:
\begin{equation}
	\label{eq:3}
	\boldsymbol{w}^{\text{cu}}(l)= \sum_{m=1}^M \frac{ |\cup_{k \in \Lambda_m} \mathcal{D}_k| }{  |\cup_{j=1:K} \mathcal{D}_j| }  \boldsymbol{w}^{\text{mec}}_m(l)
\end{equation}
Finally, the aggregated weights are broadcasted to all devices and this process is iterated until reaching a targeted convergence rate.\\
Because of the distribution and federation of the learning, a deviation between the FL and aggregated weights may occur. This divergence is studied in~\cite{hfl4}  and quantified as follows:
\begin{equation}
	\label{eq:4}
	\theta=\sum_{m=1}^{M}  \frac{ |\cup_{k \in \Lambda_m} \mathcal{D}_k| }{  |\cup_{j=1:K} \mathcal{D}_j| } \times\left\|D^{(m)}\right\|_{1},
\end{equation}
where $D^{(m)}=\left\{\left\|p(i)-p^{(m)}(i)\right\|\right\}_{i=1}^{C}$, and $p^{(m)}(i)$ is the probability of the class $i$ per MEC $m$’s distribution.

\subsection{Communication Model}
We assume that the channel between the CU and the MEC $m$ undergoes a Rician fading represented by $\mathbf{H}_m \in \mathbb{C}^{M_s \times N}$. On the other hand, we consider a Gaussian i.i.d. small-scale fading channel between the MEC $m$ and each device $k \in \Lambda_m$, denoted as $\mathbf{f}_{m,k}$. The access link between the MEC $m$ and each device $k \in \Lambda_m$ is denoted by a vector $\mathbf{g}_{m,k}= \mathbf{\beta}^{\frac{1}{2}}_{m,k} \mathbf{f}_{m,k} \in \mathbb{C}^{N_{\text{mec}} \times 1}$, where $\beta_{m,k}$ represents the pathloss. To this end, the channel matrix is defined as $\mathbf{G}_m =[\mathbf{g}_{m,k}]_{k=1:K_m} \in \mathbb{C}^{N_{\text{mec}}  \times K_m}$.\\ 
In our system, we assume a perfect Channel State Information (CSI) of all devices at each MEC. Thus, all MECs can apply a cooperative Zero Forcing (ZF) scheme to suppress the effect of inter-cell and intra-cell interferences as done in~\cite{zf}. Let ~$\mathbf{Z}_m= \mathbf{G}_m (\mathbf{G}^H_m \mathbf{G}_m)^{-1}$  be the decoder matrix of MEC $m$. Accordingly, the rate of a device $k$ connected to the MEC $m$ can be formulated as:
\begin{equation}
	\label{eq:5}
	r^{\text{dev}}_{m,k}(l)= \lambda \cdot \log_2\left(1+  \frac{ p^{\text{dev}}_{m,k}(l)}  { \sigma^2 \left[(\mathbf{G}^H_m \mathbf{G}_m)^{-1}\right]_{k,k}}\right ),
\end{equation}
where $\lambda$ is the bandwidth, and $\sigma^2$ represents the variance of an Additive White Gaussian Noise (AWGN).
 
When the MECs receive the FL weights from their related devices, they proceed the aggregation and forward their aggregated weights to the central unit. At the central unit side, in case of multiple antennas receivers, the Block Diagonalization (BD) is used to mitigate the multi-device interference. This approach is known by its high performance in MIMO systems~\cite{bd}. Therefore, we assign for each MEC $m$ a detection matrix $\mathbf{W}_m$, in order to distinguish different signals received from different MECs.
Following~\cite{bd}, $\mathbf{W}_m$ lies in the null space of matrix $\overline{\mathbf{H}}_m$, which is given by $\overline{\mathbf{H}}_m=[\mathbf{H}^T_1 ... \mathbf{H}^T_{m-1} \mathbf{H}^T_{m+1} ...\mathbf{H}^T_{M}]^T$. The decoder matrix for MEC $m$ is obtained as~$\mathbf{W}_m=\mathbf{V}^{(0)}_m \mathbf{V}^{(1)}_m$, where $\mathbf{V}^{(0)}_m$ is composed of the last $N_{\text{cu}}-\overline{J}_m$ right singular vectors of $\overline{\mathbf{H}}_m$, $\mathbf{V}^{(1)}_m$ represents the first $J_m$ right singular vectors of $\overline{\mathbf{H}}_m \mathbf{V}^{(0)}_m $, $\overline{J}_m=\text{rank}(\overline{\mathbf{H}}_m)$, and $J_m=\text{rank}(\overline{\mathbf{H}}_m \mathbf{V}^{(0)}_m)$. Hence, the uplink rate of MEC $m$ is given as~\cite{bd2}:
\begin{equation}
	\label{eq:6}
	r^{\text{mec}}_m(l)= \lambda \cdot  \log_2\left(1+ \frac{p^{\text{mec}}_m(l)  \left\|  \mathbf{H}_m \mathbf{W}_m \right\|^2_{F} }{\sigma^2}   \right).
\end{equation}

\subsection{Energy Model}
In this section, we evaluate the energy needed by the MECs and their related devices to compute different FL operations. First, the energy required by a device $k$ connected to an MEC $m$ to complete the FL tasks in the frame $l$ is expressed by: 
\begin{equation}
	\label{eq:7}
	E^{\text{dev}}_{m,k}(l)=E^{\text{cir}}+ E^{\text{cmp}}_k + E^{\text{ac}}_{m,k}(l),
\end{equation}
where $E^{\text{cir}}$ is a fixed circuit energy consumed by filters, mixers, and digital to analog converters. 
$E^{\text{cmp}}_k$ is the energy required by the device $k$ to train its local data, while $E^{\text{ac}}_{m,k}(l)$ is the energy required to transmit the resulted weights at frame $l$.\\
The energy needed by the device $k$ to train its data can be calculated as done in~\cite{cfl}:

\begin{equation}
	\label{eq:8}
	E^{\text{cmp}}_k=\varsigma \omega \vartheta^{2} Q,
\end{equation}
Where $\varsigma$ is an energy coefficient that depends on the device chip, $\omega$ represents the number of CPU cycles, $\vartheta$ is the CPU clock frequency, and Q denotes the size of the local model that highly depends on the deployed FL algorithm. It is worth mentioning that all local and global models in all devices, MECs, and CU present the same weights’ size.\\
Next, to send the weights of the local trained model to its assigned MEC $m$ at a frame $l$, the device $k$  requires an energy equal to:
\begin{equation}
	\label{eq:9}
	E^{\text{ac}}_{m,k}(l)= \frac{p^{\text{dev}}_{m,k}(l) Q }{ r^{\text{dev}}_{m,k}(l) }.
\end{equation}
where $p^{\text{dev}}_{m,k}(l)$ is the required transmit power of the device $k$ connected to the MEC $m$ at a frame $t$. Based on equation (\ref{eq:5}), $p^{\text{dev}}_{m,k}(l)$ can be formulated as:   
\begin{equation}
	\label{eq:10}
	p^{\text{dev}}_{m,k}(l)= \sigma^2 \left[(\mathbf{G}^H_m \mathbf{G}_m)^{-1}\right]_{k,k} 
	\left(2^{\frac{r^{\text{dev}}_{m,k}(l)}{\lambda}} -1 \right).            
\end{equation}
Hence, by replacing $p^{\text{dev}}_{m,k}(l)$ by its expression in (\ref{eq:9}), we obtain:
\begin{equation}
	\label{eq:11}
	E^{\text{ac}}_{m,k}(l)= \frac{ \sigma^2 Q \left[(\mathbf{G}^H_m \mathbf{G}_m)^{-1}\right]_{k,k} \left(2^{\frac{r^{\text{dev}}_{m,k}(l)}{\lambda}} -1\right)}{ r^{\text{dev}}_{m,k}(l) }.
\end{equation}

Each device in our FL system has a battery with a maximum capacity equal to $B_{max}$. These on-device batteries are also used to store the wireless transferred energy. Let $A_{m,k}(l)$ represent the amount of energy transmitted to a device $k$ connected to the MEC $m$ at a frame of time $l$. Hence, we define the battery level update of any device $k$ related to the MEC $m$ as follows:
\begin{equation}
	\label{eq:12}
	B_k(l+1)=\min \left(B_{\max}, B_k(l)-E^{\text{dev}}_{m,k}(l)+ A_{m,k}(l) \right).
\end{equation}
Furthermore, we define the energy consumed by any MEC $m$ using WET to supply its associated devices at a frame $l$ as follows: 
\begin{equation}
	\label{eq:13}
	E^{\text{mec}}_m(l)=E^{\text{cir}}+ E^{\text{wet}}_m(l) + E^{\text{bh}}_m(l),
\end{equation}
where $E^{\text{wet}}_m(l)$ denotes the energy transferred by the MEC $m$ to its related devices using the WET system, at a frame $l$. Meanwhile, $E^{\text{bh}}_m(l)$ represents the energy consumed  by the MEC $m$ to transmit the aggregated FL weights to the CU. We remind that $E^{\text{cir}}$ is a fixed circuit energy consumption. \\
Next, we adopt an optimal beamformer for a MIMO WET system to calculate the energy supplied to each device $k$ related to  MEC $m$ at a frame $l$. This energy is expressed as:

\begin{equation}
	\label{eq:14}
	A_{m,k}(l)=  E^{\text{wet}}_m(l) \beta_{m,k} \xi_{m,k} N_{\text{mec}},
\end{equation}
where $\xi_{m,k}$ is the beamforming coefficient for device $k$. \\
The energy required by the MEC $m$ to transmit the aggregated FL weights to the central unit at a frame $l$ is given by:
\begin{equation}
	\label{eq:15}
	E^{\text{bh}}_m(l)= \frac{p^{\text{mec}}_m(l) Q}{ r^{\text{mec}}_m(l) }.
\end{equation}
The transmit power $p^{\text{mec}}_m(l)$ can be derived from (\ref{eq:6}) as:
\begin{equation}
	\label{eq:16}
	p^{\text{mec}}_m(l)=\frac{\sigma^2}{\left\|  \mathbf{H}_m \mathbf{W}_m \right\|^2_{F}}  \left(2^{\frac{r^{\text{mec}}_{m}(l)}{\lambda}} -1 \right).                 
\end{equation}
Hence, by replacing $p^{\text{mec}}_{m}(l)$ by its expression in (\ref{eq:15}), we obtain:
\begin{equation}
	\label{eq:16}
	E^{\text{bh}}_m(l)= \frac{\sigma^2 Q \left(2^{\frac{r^{\text{mec}}_{m}(l)}{\lambda}} -1 \right)  }{ \left\|  \mathbf{H}_m \mathbf{W}_m \right\|^2_{F} r^{\text{mec}}_m(l) }.
\end{equation}
We assume that the energy consumed by each MEC $m$ at a frame of time $l$ is weighted by a parameter $\alpha_{m}(l)$~\cite{price}. Thus, the total cost of the energy consumed by all MECs is calculated as follows: 
\begin{equation}
	\label{eq:17}
	\Delta= \sum_{l=1}^{L} \sum_{m=1}^{M} \alpha_{m}(l) E^{\text{mec}}_m(l).
\end{equation}

\section{Problem Formulation}
Our objective in this work is to design an energy-efficient HFL framework for heterogeneous networks, where devices are supplied by wireless energy transfer. More specifically, we aim at minimizing the cost of the total grid energy consumed by different MECs by optimizing the device association and the wireless energy, constrained by the divergence between the FL’s weights and central weights. Since the deviation between the FL and central weights should not exceed a given threshold, the local data distributions at different devices should be considered when deriving the device association. On these bases, the optimization problem is formulated as (\ref{eq:20}a).\\
Constraint (\ref{eq:20}a) ensures that the deviation between the central weights and the FL’s weights does not exceed a threshold $\theta_{\max}$. Constraint (\ref{eq:20}b) guarantees that each device in the network is assigned to only one MEC. Equation (\ref{eq:20}c) ensures that the consumed energy by the devices at each frame is constrained by the available energy at the batteries. Equation (\ref{eq:20}d) ensures that we cannot store energy more than the maximum battery capacity. Finally, the last two constraints (\ref{eq:20}e) and (\ref{eq:20}f) guarantee that the transferred amounts of energy via wireless links are positive, and the device association decisions are binaries.

\begin{equation}
	\label{eq:20}
	\begin{split}
		\underset{ \underset{m=1,\ldots,M,k=1,\ldots,K,l=1,\ldots,L} {\{  \chi_{m,k}, E^{\text{wet}}_m(l) \}}  } {\min}  \Delta= \sum_{l=1}^{L} \sum_{m=1}^{M} \alpha_m(l) E^{\text{mec}}_m(l)
	\end{split}
\end{equation}
\vspace{-0.3cm}
\begin{align*}\label{c1}
	\setlength{\abovedisplayskip}{-20 pt}
	\setlength{\belowdisplayskip}{-20 pt}
	&\!\!\!\!\!\!\!\!\rm{s.\;t.}\;\scalebox{1}{$\theta \leq \theta_{\max},$}\tag{\theequation a}\\
	&\scalebox{1}{$\;\;\;  \sum_{m=1}^{M} \chi_{m,k}=1,\;\;\forall k=1,\ldots,K, $} \tag{\theequation b}\\
	&\scalebox{1}{$\;\;\; \chi_{m,k} \sum_{i=1}^{l}   E^{\text{dev}}_{m,k}(i) \leq B_k(0) + \sum_{i=1}^{l} A_{m,k}(l),$}\\
	&\scalebox{1}{$\;\;\;\forall m=1,\ldots,M,k=1,\ldots,K,l=1,\ldots,L,$} \tag{\theequation c}\\	
	&\scalebox{1}{$\;\;\; \sum_{i=1}^{l} A_{m,k}(l) - \chi_{m,k} \sum_{i=1}^{l-1} E^{\text{dev}}_{m,k}(i) \leq B_{\max},$}\\
	&\scalebox{1}{$\;\;\;\forall m=1,\ldots,M,k=1,\ldots,K,l=2,\ldots,L,$} \tag{\theequation d}\\	
	&\scalebox{1}{$\;\;\;  E^{\text{wet}}_m(l) \geq 0,\;\;\forall m=1,\ldots,M,l=1,\ldots,L, $} \tag{\theequation e}\\
	&\scalebox{1}{$\;\;\;  \chi_{m,k} \in \{0,1\},\;\;\forall m=1,\ldots,M,k=1,\ldots,K. $} \tag{\theequation f}\\
\end{align*}

Furthermore, we investigate an extended version of the formulated problem (\ref{eq:20}) by incorporating device scheduling. Some devices with high amount of wireless transferred energy may be unscheduled for certain communication rounds in order to improve the energy-efficiency of the network without deteriorating the FL training loss. The optimization problem becomes more challenging. Moreover, we extend the formulated problem (\ref{eq:20}) by incorporating a mobility scenario and assuming dynamic device association where the association index became variable over time as $\chi_{m,k}(l)$.

\section{Resource Management}
\subsection{Optimal Solution}
In this section, we derive the optimal solution of the problem formulated in (\ref{eq:20}). First of all, we can see that our optimization is a Mixed-Integer Non-Linear Programming (MINLP) due to its combinatorial nature and the non-linearity of the constraint (\ref{eq:20}a). Hence, we propose to use the following theorem to find the optimal energy transferred by each MEC at each time frame, and consequently simplify the problem.\\
\textbf{Theorem~1.} For a MEC $m$ and the set $\Lambda_m$ of its associated devices, the optimal transferred energy $E^{\text{wet*}}_m(l)$ from MEC $m$ to its associated devices is given by:
\begin{equation}
	\label{eq:21}
	E^{\text{wet*}}_m(l)=\max_{k \in \Lambda_m} \frac{E^{\text{dev}}_{m,k}(l) -B_k(l)}{\beta_{m,k} \xi_{m,k} N_{\text{mec}}}.
\end{equation}
\begin{proof}
At a frame $l$, each device $k$ related to a MEC $m$ is supplied by the harvested energy via WET, in addition to its own battery. The total energy of $k$ is expressed as follows: 
\begin{equation}
 \label{eq:22}
  E^{\text{dev}}_{m,k}(l)=B_k(l)+A_{m,k}(l).
\end{equation}
Thus, by replacing $ A_{m,k}(l)$ by its expression presented in  (\ref{eq:14}), we can derive the energy $E^{\text{wet}}_m(l)$ transferred by the MEC $m$ to  device $k$:
\begin{equation}
\label{eq:23}
E^{\text{wet}}_m(l)=\frac{E^{\text{dev}}_{m,k}(l) -B_k(l)}{\beta_{m,k} \xi_{m,k} N_{\text{mec}}}.
\end{equation}
Each MEC serves its related devices based on the highest required energy. More precisely, the MEC finds the device that requires the highest amount of energy and serves it along with the others with this same amount.
\end{proof}
Using the expression of the transferred energy $E^{\text{wet}}_m(l)$ presented in  equation (\ref{eq:21}), our optimization problem (\ref{eq:20}) can be simplified  as it becomes an integer non-linear problem. This simplified problem that aims now to find the optimal device association is combinatorial and is known to be NP-hard~\cite{np}. However, it can be solved using Brute-Force Search (BFS). Although deriving the optimal solution is still highly complex, the results can be used as a benchmark to our heuristic approach that will be proposed in what follows. It is also worth mentioning that by neglecting the constraint (\ref{eq:20}a), the optimization becomes an integer linear problem that can be solved using numerical tools such as CVX.

\subsection{Heuristic Approach}
\begin{figure}[h]
	\centerline{\includegraphics [width=0.95\columnwidth]  {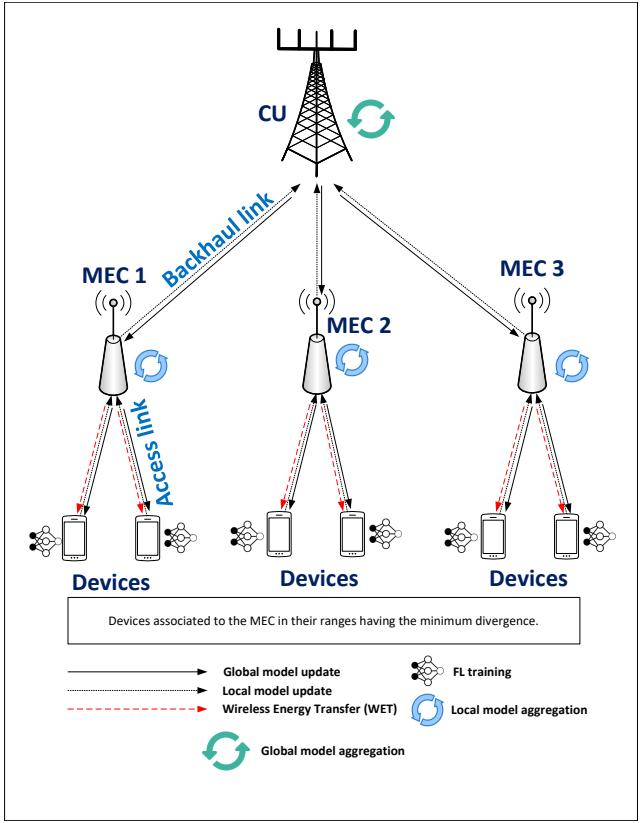}}
	\caption{Diagram of the proposed algorithm H2RMA.}
	\label{fig1a}
\end{figure}

\begin{figure}[h]
	\centerline{\includegraphics [width=0.95\columnwidth]  {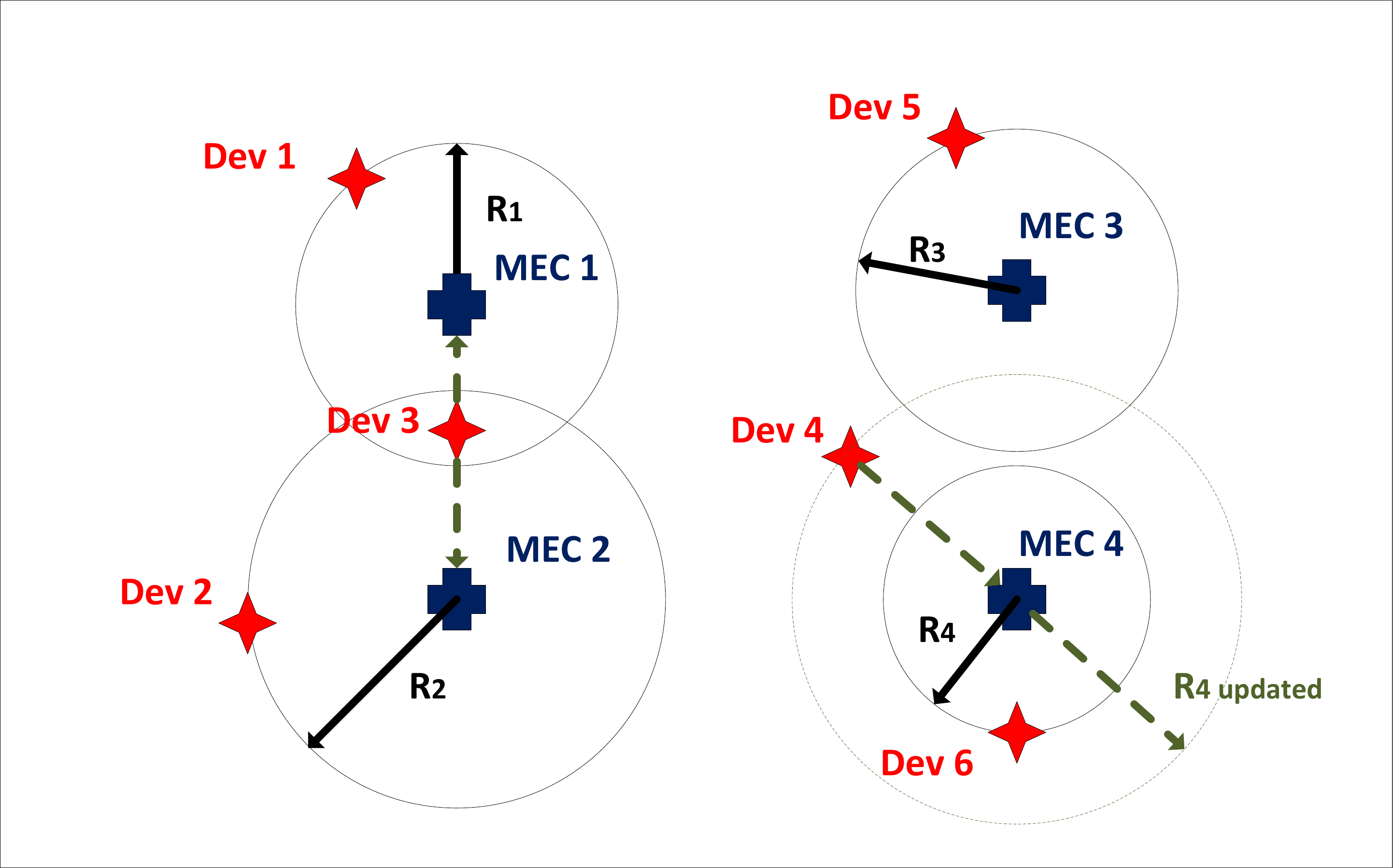}}
	\caption{An example of the device association procedure.}
	\label{fig2a}
\end{figure}

Due to the combinatorial complexity of the optimal solution that may prohibit its online deployment, we propose a heuristic algorithm that solves the device association and energy management problem with low complexity. Similarly to the optimization, our algorithm aims to minimize the cost of the consumed grid energy, while respecting the FL divergence constraints.\\
Let $R_m$ denote the range of the MEC $m$, which is the pathloss of the farthest associated device. Based on (\ref{eq:15}), we can see that the device with the higher pathloss receives the lowest amount of energy via WET. In other words, any device assigned to the MEC $m$ with a lower pathloss than $R_m$, i.e., $\beta_{m,k} \leq R_m$, may benefit from a higher transferred energy. Accordingly, each device $k$ tries to find the set of MECs having a higher range than its pathloss. Next, it selects the one with minimum divergence $\theta$. This device association process is done one by one. If one of the devices fails to find a MEC with a suitable range, it will be associated to the closest one. Our proposed device association heuristic reduces not only the energy consumption but also the divergence between FL and central weights, which contributes to minimize the training loss. Finally, for each frame $l$, we can calculate the amount of energy transmitted by each MEC via WET using equation (\ref{eq:21}). Algorithm~1 presents different steps of our proposed HFL Heuristic Resource Management Algorithm (H2RMA). Moreover, the proposed algorithm is described in Fig.~\ref{fig1a} and we provide an example of the device association procedure in Fig.~\ref{fig2a}. We can see in Fig.~\ref{fig2a} that Dev~3 is within the range of MEC~1 and MEC~2. Hence, it chooses to be associated with the MEC that minimizes the divergence. For Dev~4, it is outside the range of all the MECs. Hence, it chooses to be associated with the MEC under minimal energy requirement. Then, the range of MEC~4 is updated following Dev~4.

We investigate the complexity of H2RMA. The device association process is presented in lines (3-20) using two loops \textbf{for} with a complexity equal to $O(M K)$. In the \textbf{for} loop in lines (21-29), the power control is conducted with a complexity in the order $O(L(2K+M))$. Hence, the complexity of H2RMA can be expressed as:
\begin{equation}
	\label{eq:24}
	C^{\text{H2RMA}}= O(M K + L(2K+M)).
\end{equation}
Hence, H2RMA can be executed in polynomial time.

\begin{algorithm}[h]
	\begin{algorithmic}[1]
		\caption{HFL Heuristic Resource Management Algorithm (H2RMA)}	\State \textbf{Initialization:}
		\State $\chi_{i_k,k}\gets 0,~m=1..M,~k=1..K$, 
		\State $R_m \gets 0,~m=1..M$
		\For{$k=1:K$}
		\State Initialize $i_k \gets 0$
		\State Initialize $\theta_{\min} \gets \theta_{\max}$
		\For{$m=1:M$}
		\If{$\beta_{m,k} \leq R_m$}, verify if device $k$ is within the range of MEC $m$.
		\State compute $\theta$
		\If{$\theta \leq \theta_{\min}$}
		\State $i_k \gets m$, select the MEC with the minimum divergence.
		\State $\theta_{\min} \gets \theta$ 
		\EndIf
		\EndIf
		\EndFor
		\If{$ i_k  = 0$}, if device $k$ is not within the range of any MEC.
		\State $i_k \gets \argmin_{m=1..M} \beta_{m,k} $, associate with the closest MEC $m*$.
		\State $R_{m*} \gets \beta_{m*,k}$, update the range of the MEC $m*$.
		\EndIf
		\State $\chi_{i_k,k} \gets 1$     
		\EndFor
		\For{$l=1:L$}
		\State compute the energy required by each device.
		\State compute $E^{\text{wet}}_m(l),~m=1..M$ using (\ref{eq:21}).
		\State send the weights $\boldsymbol{w}_k(l)$ to the related MECs and aggregation using (\ref{eq:3}).
		\State $B_k(l+1) \gets \min \left(B_{\max}, B_k(l)-E^{\text{dev}}_{m,k}(l)+ A_{m,k}(l) \right)$, battery update
		\State send the weights $\boldsymbol{w}^{\text{mec}}_m(l)$ to the central unit to compute the aggregation using (\ref{eq:4}).    
		\State send the aggregated weights to all devices.
		\State different devices update their FL training models.
		\EndFor
	\end{algorithmic}
\end{algorithm}

\subsection{Device Scheduling}
We now investigate the device scheduling for our proposed HFL framework. In fact, at a given frame $l$, some devices may face bad channel conditions and require high amount of wireless transferred energy, without contributing much on the HFL procedure. This costs the system unnecessary energy consumption that can be avoided. In this scenario, a practical idea to save the grid energy is to deactivate some devices for some frames based on the available energy and time-varying channel. Moreover, the scheduled devices at each frame should respect the maximal divergence constraint so as not to degrade of FL performance.\\
Including device scheduling to the optimization problem (\ref{eq:20}) which is MINLP increases the complexity of finding the optimal solution. This optimal device scheduling can be obtained with BFS. However, its high computational complexity prevents it from being conceivable. Hence, we propose an efficient heuristic device scheduling scheme to save unnecessary grid energy cost. We start by considering that all devices are selected. The devices to be deactivated at the current frame are those that require high amount of energy exceeding a certain threshold $E_{\text{th}}$ given by the set:
\begin{equation}
	\label{eq:25}
	\Omega=\{k \in \{1,\ldots,K\} | E^{\text{ac}}_{m,k}(l) > E_{\text{th}}\} .
\end{equation}
We deactivate the devices in $\Omega$ one by one iteratively and we start by the ones that require higher transmit energy. At each iteration, the device is deactivated only if the maximal divergence constraint is satisfied so as not to degrade FL performance.

\section{Impact of Mobility}
In this section, we investigate an applied case of occupational health and safety by applying our proposed HFL framework in order to minimize the health risks of the workers in a construction site. In consequence, we choose a mobility scenario~\cite{mobility} that reflects the workers equipped with smart devices in a construction site. The dynamic interactions between workers on the site makes the construction job one of the most risky activities. The mobility of workers can be modeled by Hidden Markov Models (HMMs)~\cite{mobility}, which are powerful statistical tools that represent the probability distributions over observation. We can categorize the workers into three states, i.e., static, normal, and risky, depending on their speed and turning angle. Hence, the hidden states of the HMM may be defined depending on the values of these two parameters, i.e., speed and turning angle. These values are provided in Table~\ref{tab2}. Moreover, the step length of the workers is modeled by the Gamma distribution and the turning angle is modeled by the Von Mises distribution. Let us divide the network into micro-cells with length $\mu$. The average number of frames that a normal worker moves from a micro-cell to another is given by:
\begin{equation}
\label{eq:26}
\psi_{\text{nml}}= \lceil \frac{\mu}{T V_{\text{nml}}} \rceil,
\end{equation}
where $\lceil \cdot \rceil$ is the ceiling function, and $V_{\text{nml}}$ is the average speed of a normal worker. The average number of frames that a risky worker moves from a micro-cell to another is given by:
\begin{equation}
\label{eq:27}
\psi_{\text{rsk}}= \lceil \frac{\mu}{T V_{\text{rsk}}} \rceil,
\end{equation}
where $V_{\text{rsk}}$ is the average speed of a risky worker. It is clear that the risky workers change their micro-cells more frequently than the normal workers. Also, some workers when they pass by certain places in the network (like office, meeting room, coffee break ...), they remain static for a certain period.

\begin{table}[h]
	\centering
	\caption{Values of Hidden States.}
	\label{tab2}
	\begin{tabular}{|C{1.1cm}|C{2.6cm}|C{3.7cm}|}
		\hline  \textbf{State} & \textbf{Speed (steps/mn)} &  \textbf{Turning angle (radians)} \\
		\hline   Static & 0 & angle < $\pi/2$ \\
		\hline   Normal & 0 < steps $\leq$ 84  & $\pi/2 \leq$ angle < $\pi$ \\
		\hline   Risky & steps > 84 & angle $\geq \pi$  \\
		\hline
	\end{tabular}
\end{table}

Considering this mobility scenario, the mobile devices move around and become far from their associated MECs. Hence, they require higher amounts of wireless transferred energy. Indeed, performing fixed device association scheme when the devices are moving involves an increase in the energy consumption cost. Therefore, we need to design an adaptive device association scheme taking into account devices mobility. 

\begin{algorithm}[h]
	\begin{algorithmic}[1]
		\caption{Dynamic HFL Device Association (DHDA)}	
		\For{$l=1:L$}
		\If{$ l \mod \psi_{\text{rsk}} = 0$ or $ l \mod \psi_{\text{nrm}} = 0$ }, 
		\State Update $\Gamma$ the set of mobile devices that changed their micro-cell during the current frame.
		\For{$k \in \Gamma $}
		\State $i_k \gets 0$, initialization
		\State $\theta_{\min} \gets \theta_{\max}$, initialization
		\For{$m=1:M$}
		\If{$\beta_{m,k} \leq R_m$}, verify if the mobile device $k$ is within the range of MEC $m$.
		\State compute $\theta$
		\If{$\theta \leq \theta_{\min}$}
		\State $i_k \gets m$, select the MEC that has the minimum divergence.
		\State $\theta_{\min} \gets \theta$ 
		\EndIf
		\EndIf
		\EndFor
		\If{$ i_k  = 0$}, if the mobile device $k$ is not within the range of any MEC.
		\State $i_k \gets \argmin_{m=1..M} \beta_{m,k} $, associate with the closest MEC $m*$.
		\State $R_{m*} \gets \beta_{m*,k}$, update the range of the MEC $m*$.
		\EndIf
		\State $\chi_{i_k,k} \gets 1$     
		\EndFor		
		\EndIf
		\EndFor
	\end{algorithmic}
\end{algorithm}

We propose a heuristic dynamic device association strategy for HFL over HetNets enabled by WET. We start by executing the original device association algorithm H2RM presented in section V. Then, we change the associated MECs of the normal workers after each $\psi_{\text{nml}}$ frames and for the risky workers after each $\psi_{\text{rsk}}$ frames. The mobile device $k$ updates its new associated MEC as follows. It finds first the new set of MECs that have a range higher than its pathloss. We already defined the range $R_m$ of a MEC $m$ as the pathloss of the farthest associated device. Hence, being associated with one of these MECs will not induce any supplementary energy cost. The mobile device will choose the MEC having the minimum divergence $\theta$. 
If this device fails to find a MEC with suitable range, the closest one will be selected.
The proposed Dynamic HFL Device Association Algorithm (DHDA) is illustrated in Algorithm~2.\\ 
To find the complexity of DHDA, we need to multiply the computational complexity of the first \textbf{for} loop of H2RMA by the number of times that we repeat association of mobile device $\lceil \frac{L}{\psi_{\text{rsk}}} \rceil$. Hence, the complexity of DHDA is give by:
\begin{equation}
	\label{eq:32}
	C^{\text{DHDA}}= O( \lceil  \frac{L}{\psi_{\text{rsk}}} \rceil M K  + L(2K+M)).
\end{equation}

\section{Numerical Results}

In this section, we investigate the performance of our proposed HFL framework under different network configurations. The simulation settings are as follows: we positioned the CU at the center of a circular macro-cell and uniformly distributed the MECs and devices within the considered coverage. First, we consider that devices are required to recognize digits between 0 and 9, which can be simulated by training a classification on MNIST dataset. In our simulation, we used a deep learning model to train on the MNIST dataset. This model is composed of two convolutional layers of 3x3 filters and we chose Rectified Linear Unit (ReLU) as an activation function. To downsize the resulting feature maps, we added a max pooling layer, which output is flattened to be fed into the subsequent fully connected layer. Finally, the last layer is designed to generate the probabilities of the 10 classes. We note that we used the dropout technique~\cite{dl} to reduce the overfitting and regularize the model. Furthermore, we incorporated the information about the class weight in the training process to mitigate the problem of imbalanced classes, and we used simple Stochastic Gradient Descent (SGD) to train the model and optimize the cross-entropy loss, throughout 20 training epochs.
Next, we investigate CIFAR-10 (Canadian Institute For Advanced Research) which is a well-known image dataset that is commonly used to validate the performance of machine learning and computer vision algorithms. The data consist of 60000 color images of size 32x32 and 10 classes; each class has 6000 images. The class labels are: horse, frog, dog, cat, deer, bird, automobile, airplane, truck, and ship. The data does not have overlap between classes, i.e. the classes are mutually exclusive. We summarize the simulation parameters used in this evaluation section in Table~\ref{tab3}.

\begin{table}[h]
	\centering
	\caption{Configuration of the HFL over HetNet environment in our experiments.}
	\label{tab3}
	\begin{tabular}{|C{1cm}|L{4.1cm}|C{2cm}|}
		\hline  \textbf{Symbol} & \textbf{Description} &  \textbf{Value}  \\
		\hline   $N_{\text{cu}}$ & number of antennas at the CU & 128 \\
		\hline   $M$ & number of MECs & 8 \\
		\hline   $N_{\text{mec}}$ & number of antennas at each MEC & 16 \\
		\hline   $K$ & number of devices & 20 \\
		\hline   $L$ & number of frames & 50 \\	
		\hline   $C$ & number of classes & 10 \\		
		\hline   $B_{\textrm{max}}$ & max battery capacity & 1000~J \\
		\hline   $B_{k}(0)$ & initial battery level & 200~J \\
		\hline   $\nu$ & path loss exponent & 3.7 \\
		\hline   $\lambda$  & bandwidth & 20~MHz\\
		\hline   $\vartheta$  & frequency of the CPU clock & $10^9$ \\
		\hline   $\omega$  & number of CPU cycles & 40\\
		\hline   $\varsigma $  & consumed energy coefficient & $10^{-27}$\\
		\hline   $\theta_{\text{max}}$  & maximal divergence & 0.5 \\
		\hline    & cell radius & 200 m \\
		\hline    & circuit power per RF chain  & 30 dBm\\
		\hline    & noise power spectral density & -174 dBm/Hz \\
		
		\hline
	\end{tabular}
\end{table}

\subsection{Fixed Device Association without Scheduling}

\begin{figure}[h]
	\centerline{\includegraphics [width=1\columnwidth]  {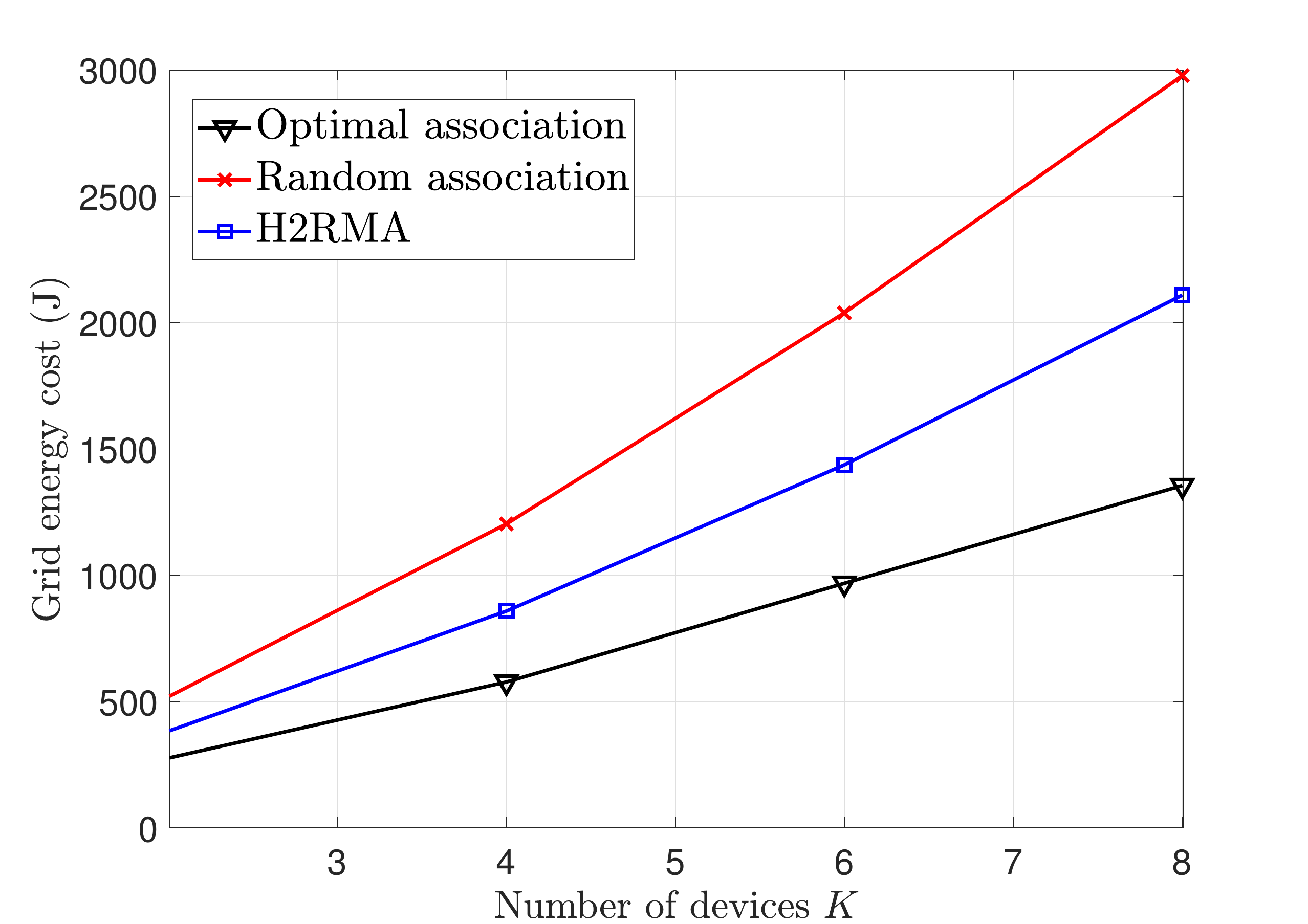}}
	\caption{Grid energy cost while varying the number of devices $K$: comparison between optimal device association, random association, and H2RMA ($M=2,B_{k}(0)=10~\text{J}$).}
	\label{fig2}
\end{figure}

\begin{figure}[h]
	\centerline{\includegraphics [width=1\columnwidth]  {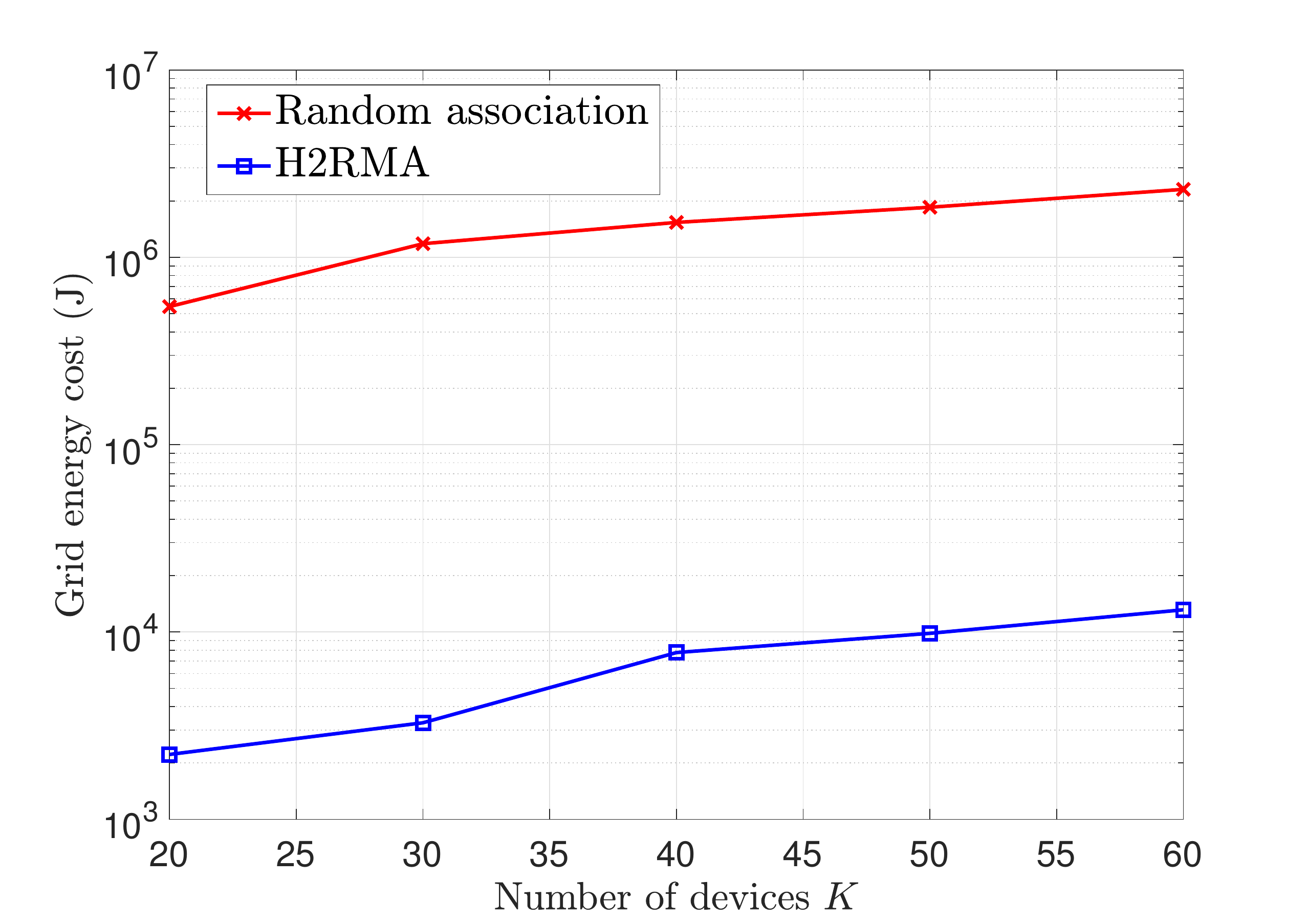}}
	\caption{Grid energy cost while varying the number of devices $K$: comparison between H2RMA and random device association scheme ($M=8,B_{k}(0)=200~\text{J}$).}
	\label{fig3}
\end{figure}

\begin{figure}[h]
	\centerline{\includegraphics [width=1\columnwidth]  {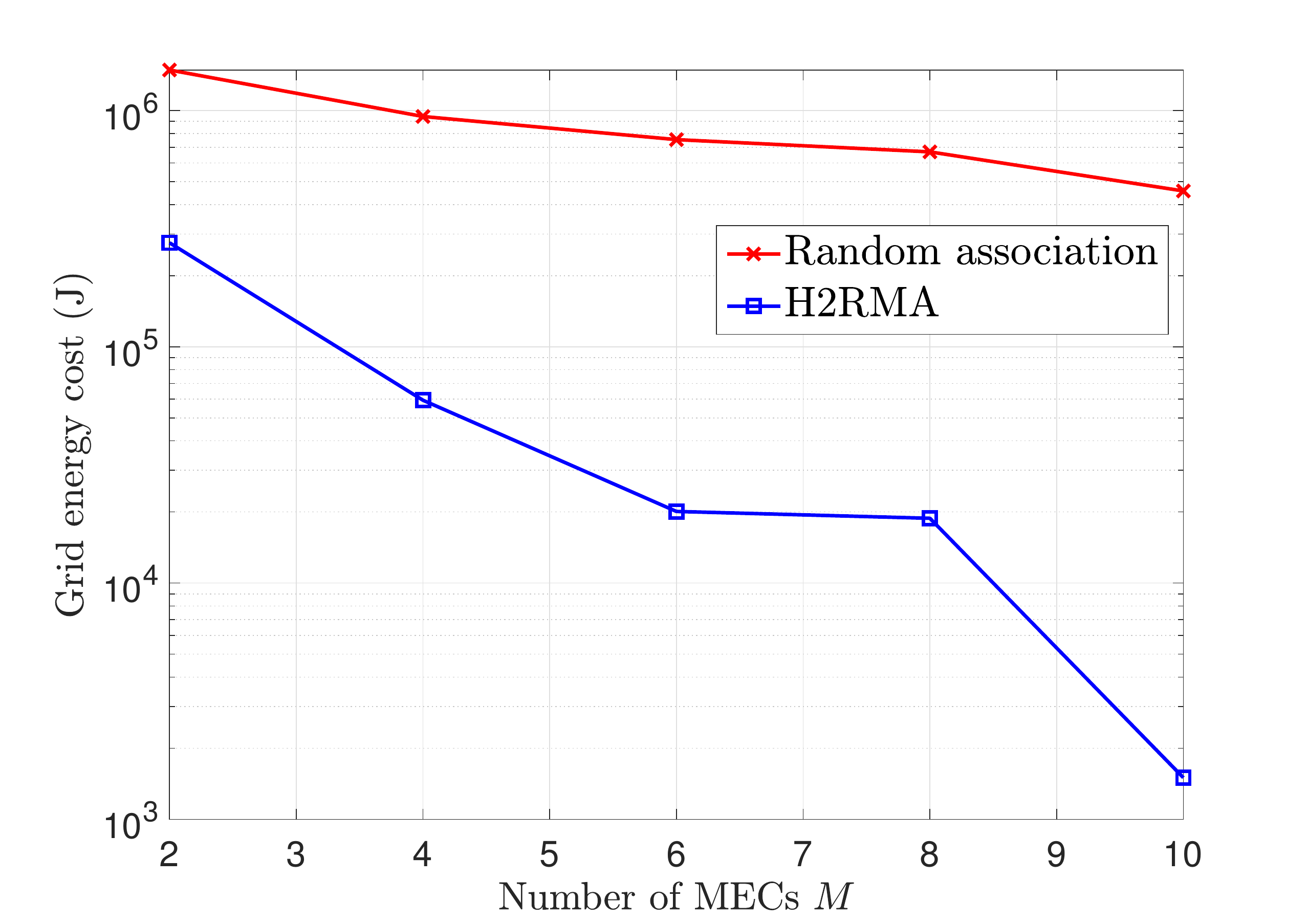}}
	\caption{Grid energy cost while varying the number of MECs $M$: comparison between H2RMA and random device association  ($K=20,B_{k}(0)=200~\text{J}$).}
	\label{fig4}
\end{figure}

We start by investigating the performance of the proposed HFL framework considering a static network and fixed device association over time. In Fig.~\ref{fig2}, we show the grid energy cost, when varying the number of devices. Deriving the optimal device association suffers from a high computation complexity. Thus, we run our simulation using only two MECs in order to establish the optimal solution as a baseline. Compared to the optimal solution, we can see that H2RMA achieves a good performance in terms of energy consumption cost, while offering a much lower computation complexity. In Fig.~\ref{fig3}, we illustrate the  grid energy cost incurred by our heuristic compared to the random device association scheme, when changing the number of devices $K$. Using a large number of devices, H2RMA presents a highly lower grid energy cost, due to its adequate device association method. Fig.~\ref{fig4} depicts the performance of our heuristic algorithm when varying the number of MECs. When the number of MECs increases, the cost of grid energy decreases, as different devices in the system are surrounded by a larger set of MECs; hence, they have a higher probability to be associated to closer MECs. Moreover, we can see clearly, that our H2RMA algorithm outperforms the random association scheme, for different numbers of MECs.

\begin{figure}[h]
	\centerline{\includegraphics [width=1\columnwidth]  {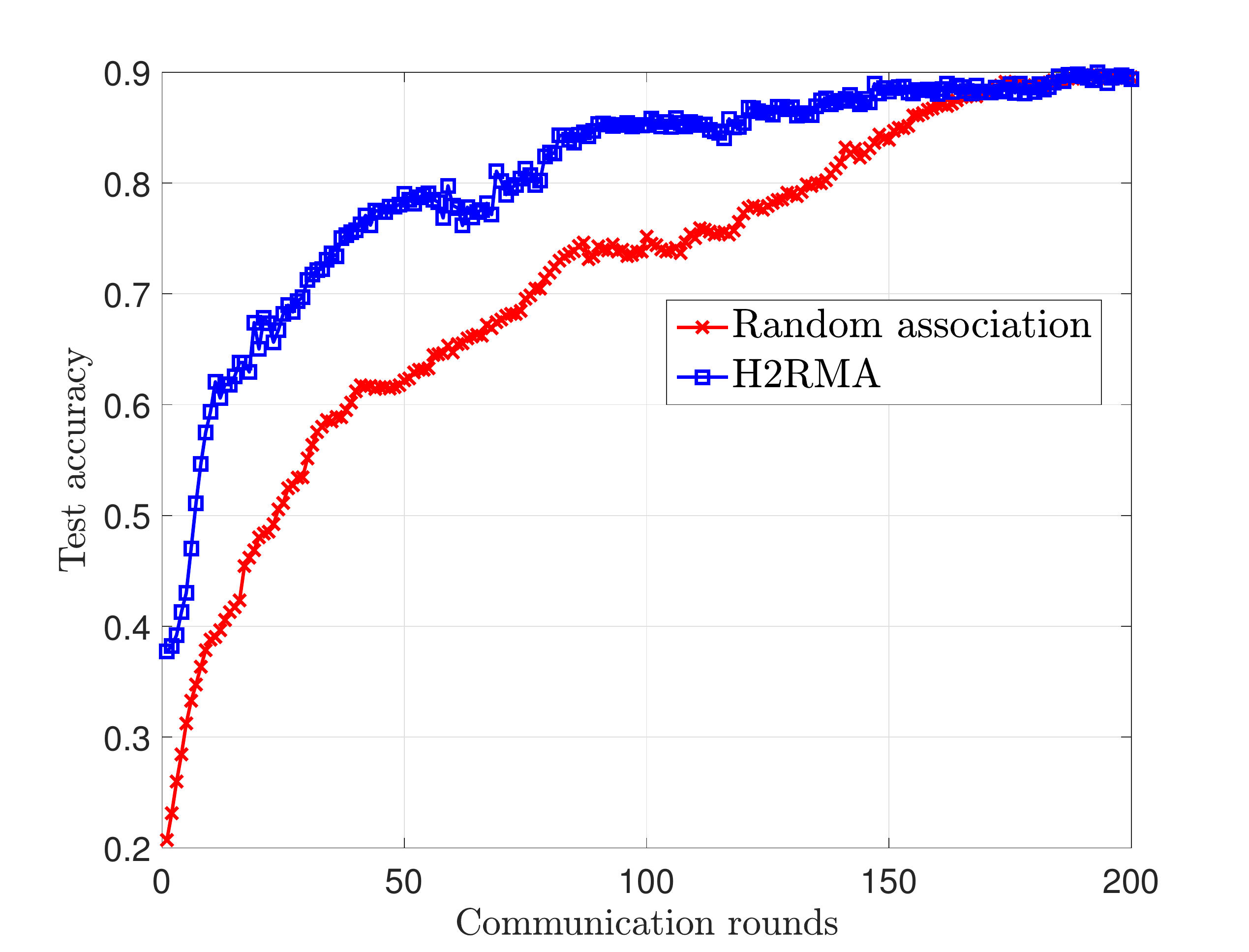}}
	\caption{Test accuracy: comparison between H2RMA and random device association scheme considering MNIST ($K=15,M=3,C=3$).}
	\label{fig5}
\end{figure}

\begin{figure}[h]
	\centerline{\includegraphics [width=1\columnwidth]  {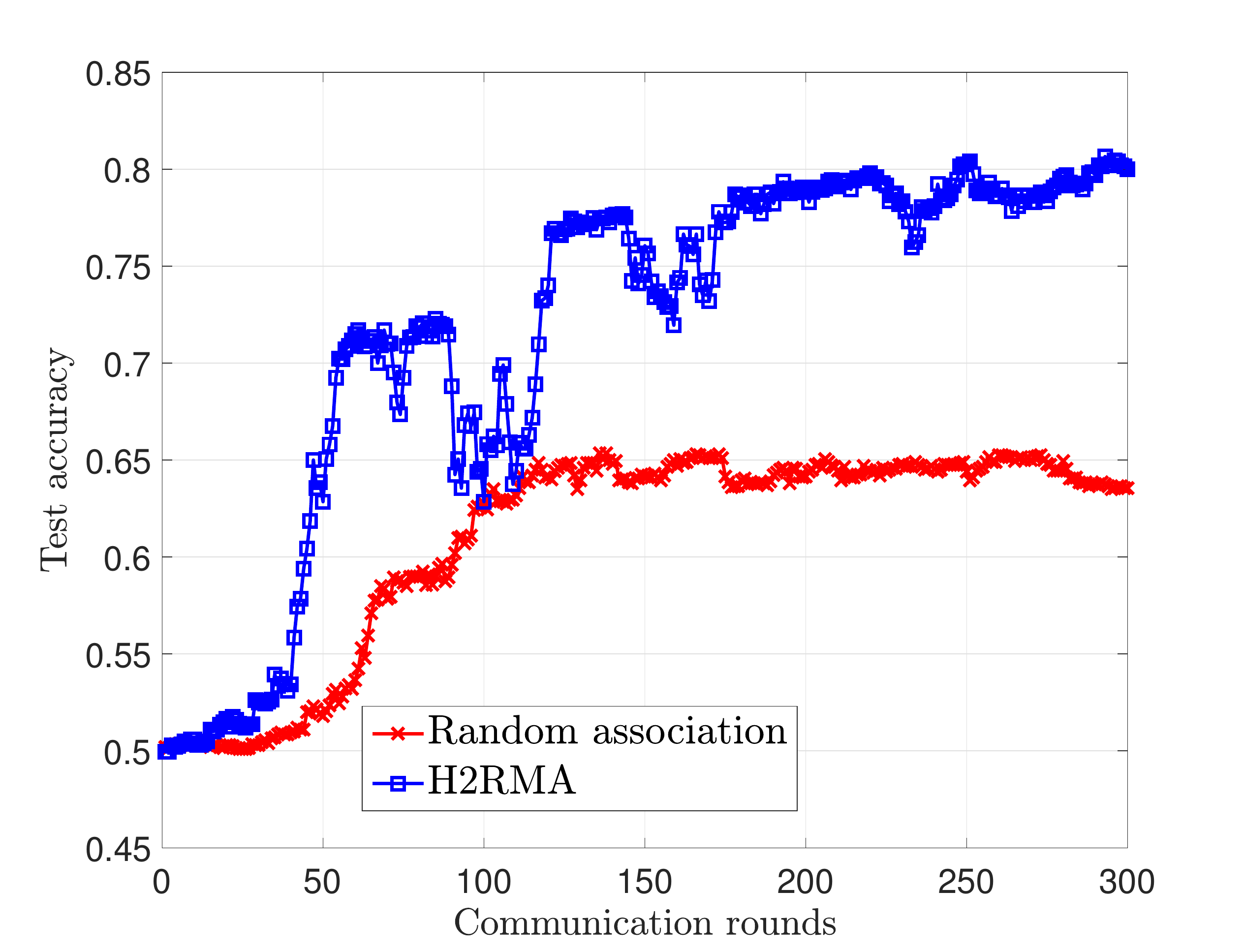}}
	\caption{Test accuracy: comparison between H2RMA and random device association scheme considering CIFAR-10 ($K=15,M=3,C=2$).}
	\label{fig6}
\end{figure}

Fig.~\ref{fig5} presents the accuracy performance of H2RMA compared to the random association scheme, when tested using MNIST dataset. We can see that our algorithm converges faster, owing to its efficient device association process that minimizes the divergence. In Fig.~\ref{fig6}, the accuracy performance of H2RMA is evaluated, while considering CIFAR-10 dataset. Similar to the MNIST dataset, our proposed device association scheme performs well and outperforms the random device association scheme in terms of test accuracy.

\subsection{Fixed Device Association with Scheduling}
\begin{figure}[h]
	\centerline{\includegraphics [width=1\columnwidth]  {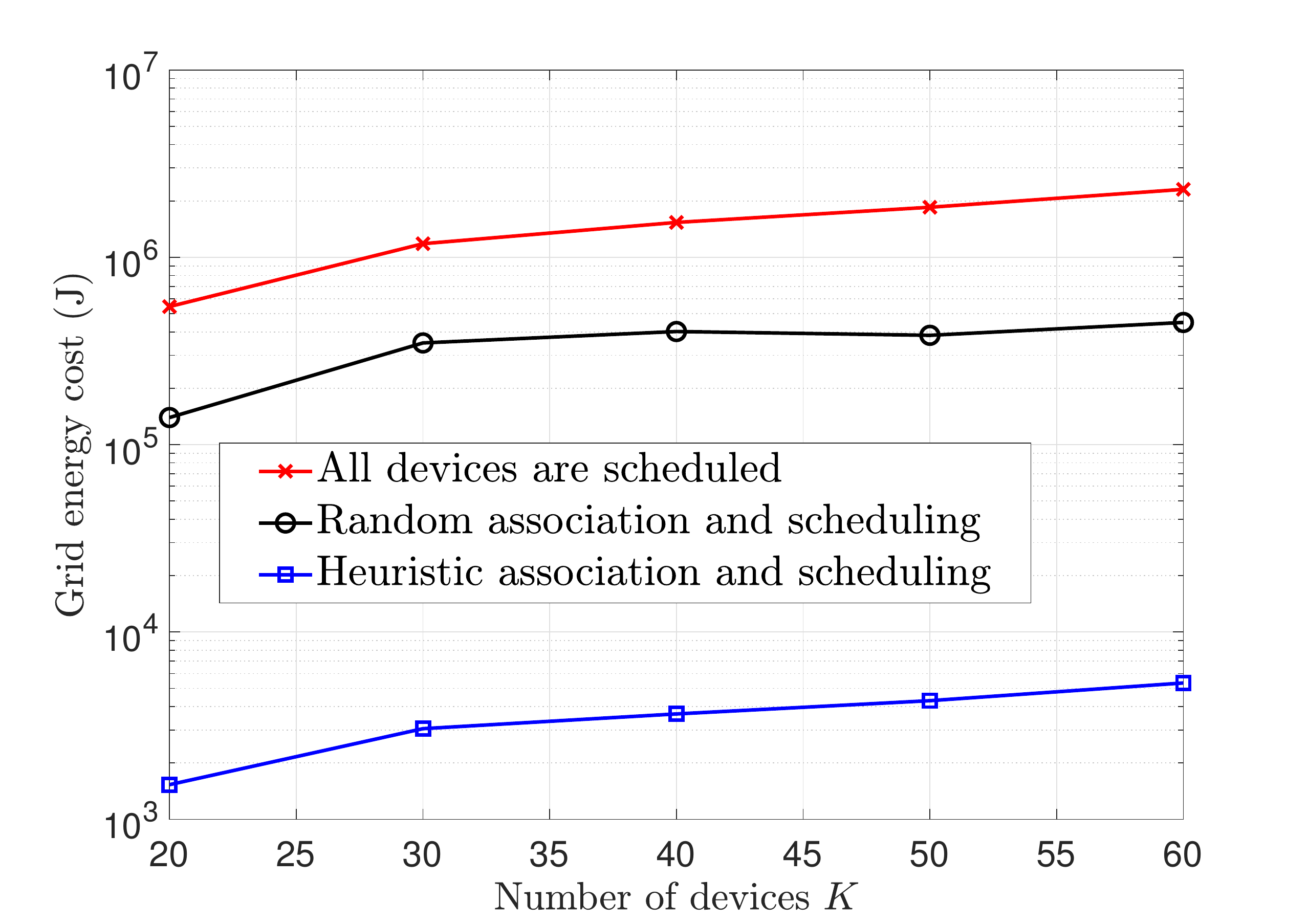}}
	\caption{Grid energy cost while varying the number of devices $K$: comparison between the proposed scheduling heuristic, random device association and system scheduling all devices ($M=8,B_{k}(0)=200~\text{J}$).}
	\label{fig7}
\end{figure}

We investigate now in Fig.~\ref{fig7} the impact of the proposed device scheduling on the grid energy consumed by our system. It is clear that we can save a significant amount of energy while deactivating some devices that require a high amount of wireless transferred energy for a certain number of frames. Moreover, the proposed device scheduling scheme performs very well compared to the random scheme, because only far devices are being deactivated, while respecting the maximal divergence constraint.

\begin{figure}[h]
	\centerline{\includegraphics [width=1\columnwidth]  {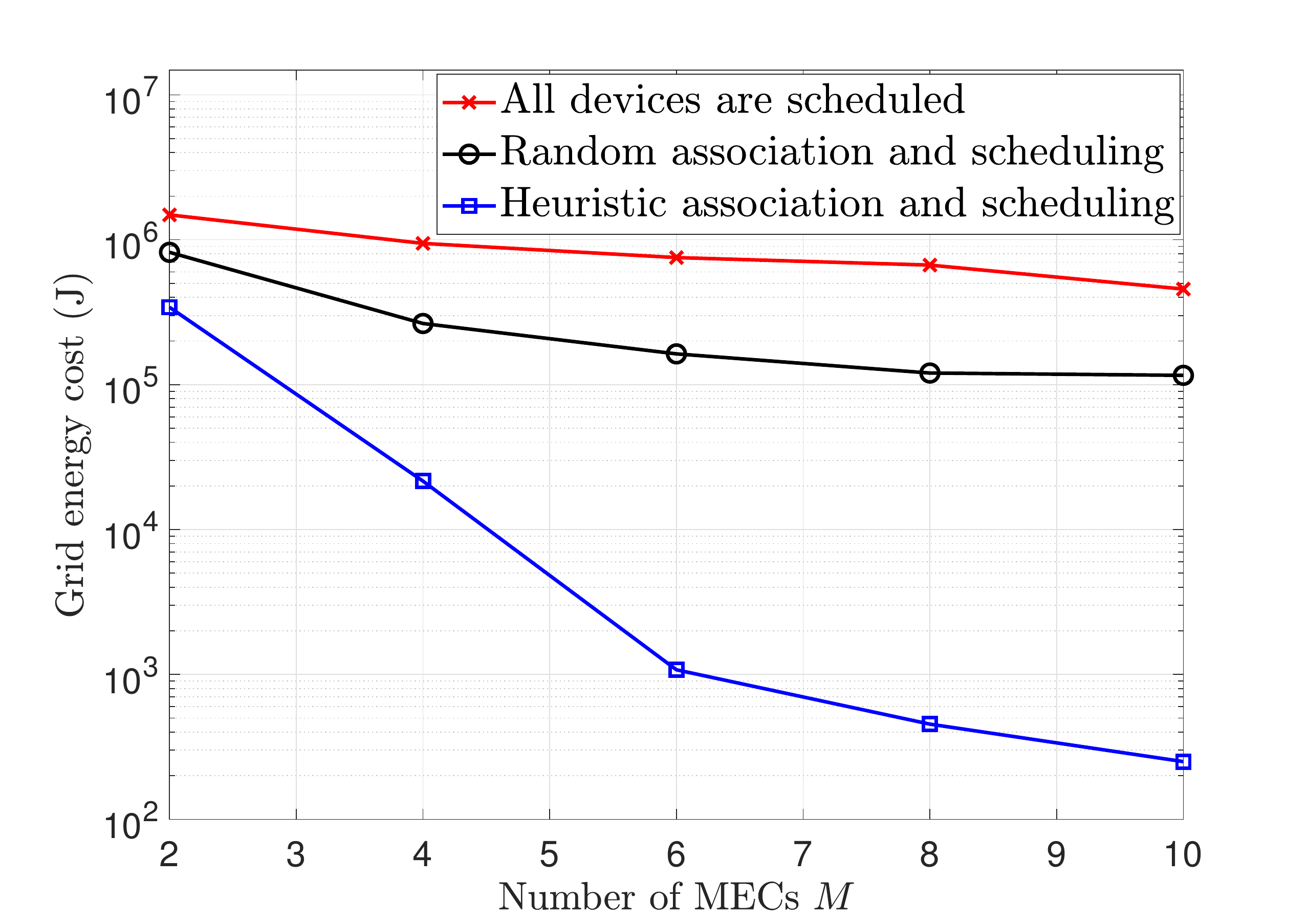}}
	\caption{Grid energy cost while varying the number of MECs $M$: comparison between the proposed scheduling heuristic, random device association and system scheduling all devices ($K=20,B_{k}(0)=200~\text{J}$).}
	\label{fig8}
\end{figure}

Fig.~\ref{fig8} depicts the performance of our device scheduling heuristic when varying the number of MECs. We can save a significant amount of grid energy specifically when the number of MECs increases thanks to the efficient device scheduling scheme.

\begin{figure}[h]
	\centerline{\includegraphics [width=1\columnwidth]  {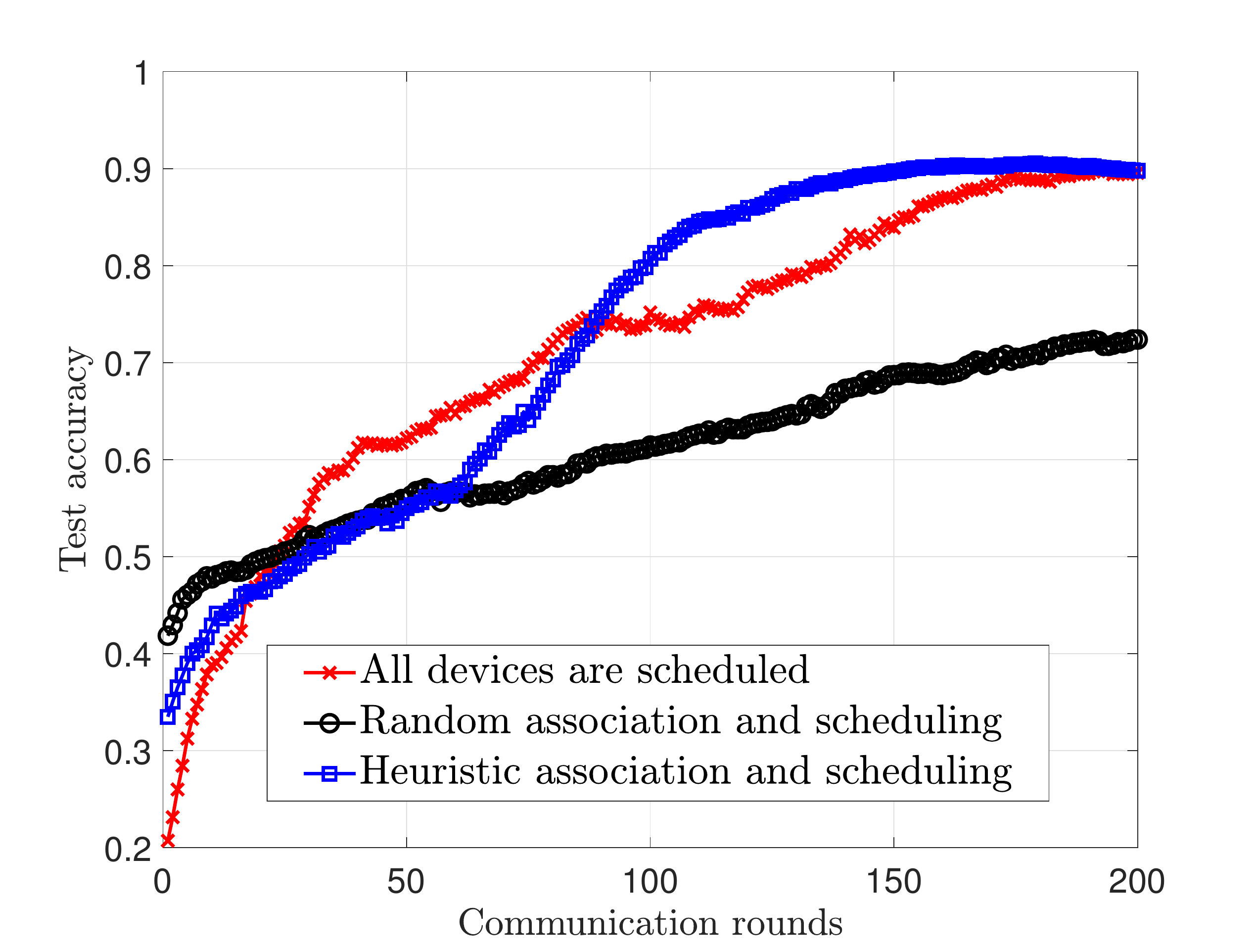}}
	\caption{Test accuracy considering MNIST: comparison between the proposed scheduling heuristic, random device association and system scheduling all devices  ($K=15,M=3,C=3$).}
	\label{fig9}
\end{figure}

Fig.~\ref{fig9} depicts the performance of our heuristic device scheduling algorithm in terms of test accuracy considering MNIST. It can be seen that when deactivating some devices randomly, the test accuracy decreases. However, the proposed heuristic device scheduling scheme allows to enhance the convergence rate since only a subset of devices that respect the divergence constraint is scheduled.

\subsection{Dynamic Device Association}
\begin{figure}[h]
	\centerline{\includegraphics [width=1\columnwidth]  {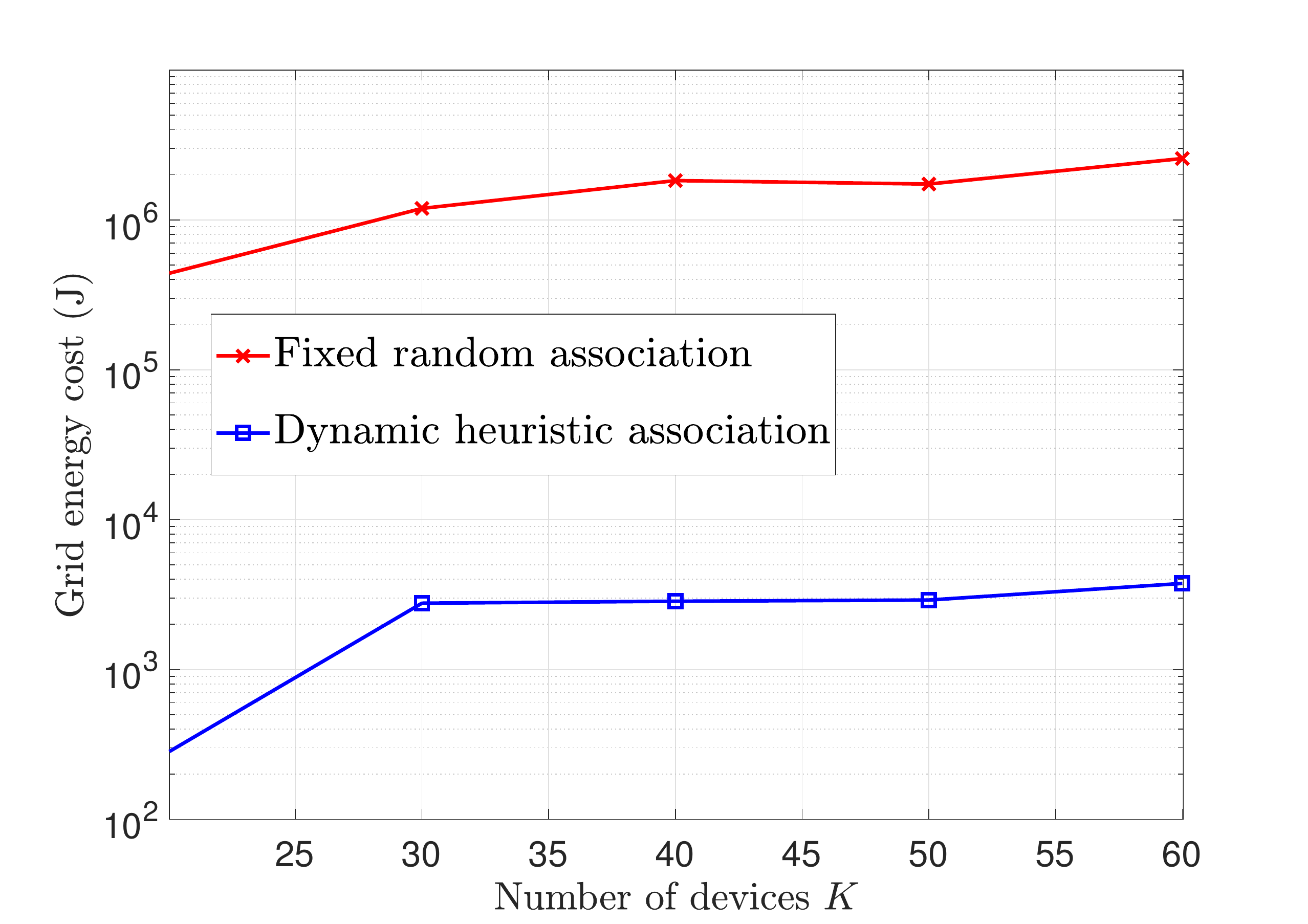}}
	\caption{Grid energy cost while varying the number of devices $K$ considering a mobility scenario ($M=8,B_{k}(0)=200~\text{J}$).}
	\label{fig10}
\end{figure}

Finally, we implement a mobility scenario for the devices which is described in Section V. The mobility parameters used in this simulation are illustrated in Table~\ref{tab2}. In Fig.~\ref{fig10}, we evaluate the cost of grid energy consumed by our HFL framework, when considering the mobility scenario. 
Indeed, the proposed dynamic device association algorithm allows saving significant amount of energy because it allows us to efficiently adapt the association of the mobile devices. Furthermore, the performance gap between dynamic and fixed device association increases when the number of mobile devices in the network becomes higher.

\begin{figure}[h]
	\centerline{\includegraphics [width=1\columnwidth]  {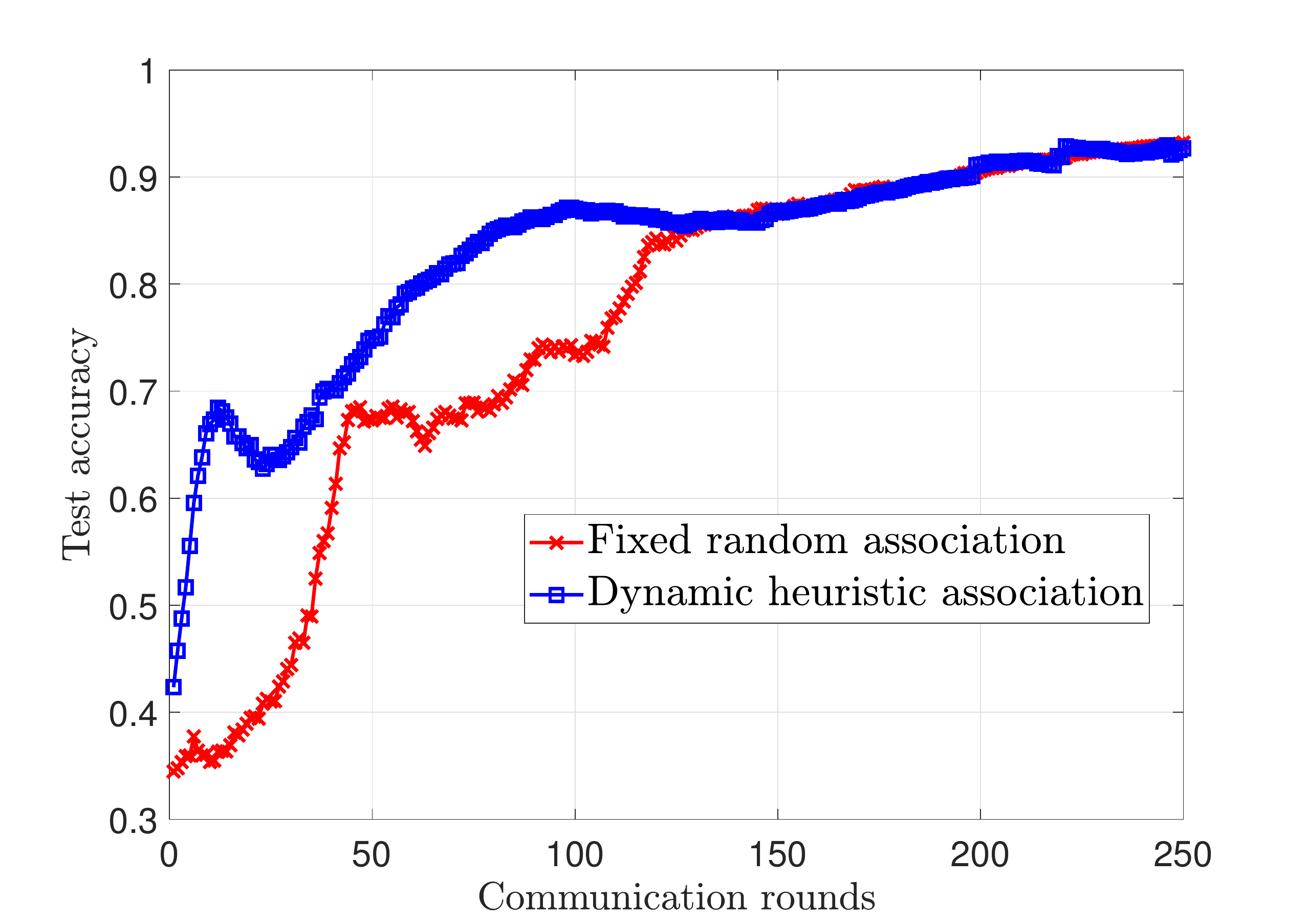}}
	\caption{Test accuracy considering a mobility scenario and MNIST dataset ($K=15,M=3,C=3$).}
	\label{fig11}
\end{figure}

Fig.~\ref{fig11} shows the performance of the proposed HFL framework in terms of convergence rate considering a mobility scenario. The proposed dynamic device association scheme ensures that mobile devices respect the maximal divergence constraints and hence achieve better convergence rate than the fixed device association.

To summarize, the performance of our HFL framework is evaluated by showing the energy-efficiency of the proposed resource management approaches, in addition to the high accuracy achieved under different network configurations and using two well-known datasets, namely MNIST and CIFAR-10. Moreover, we showed the efficiency of the proposed dynamic device association considering a mobility scenario in the network.

\section{Conclusion}
An energy-efficient HFL over HetNets with massive MIMO wireless backhaul enabled by WET has been investigated in this work. An optimization problem involving a grid energy consumption cost constrained by energy availability and maximal divergence, have been developed. In addition, an optimal energy management has been derived. To achieve this, we have formulated the main objective as a device association optimization problem that we solved using BFS. Because of the high complexity to get the optimal solution, we have proposed an efficient device association heuristic, with low computational complexity, namely H2RMA. Moreover, a device scheduling strategy has been designed to save the grid energy without deteriorating the learning accuracy. The impact of mobility on FL performance has also been investigated by developing an efficient dynamic device association scheme. Our extensive experiments confirm the efficiency of our HFL framework in terms of accuracy, convergence rate and grid energy consumption, when evaluated under different network configurations.
 
As a future work, we plan to develop an efficient clustering strategy based on reinforcement learning for HFL over HetNets enabled by WET.

\section*{Acknowledgment}
This work was supported by National Priorities Research Program-Standard (NPRP-S) Thirteen (13th) Cycle grant $\#$~NPRP13S-0205-200265 from the Qatar National Research Fund (a member of Qatar Foundation).  The findings achieved herein are solely the responsibility of the authors.

\balance
	
\end{document}